\documentclass[aps,prd,reprint,twocolumn,superscriptaddress,longbibliography,nofootinbib,floatfix,showpacs]{revtex4-2}

\usepackage{epsfig}
\usepackage{amsmath,amssymb,amsfonts}
\usepackage{hyperref}
\usepackage{mathrsfs}
\usepackage{bbm}
\usepackage{slashed}
\usepackage{graphicx}
\usepackage{verbatim}

\usepackage{bm}

\usepackage[usenames]{color}

\definecolor{darkgreen}{rgb}{0.2,0.6,0}

\newcommand{\be}{\begin{equation}}
\newcommand{\ee}{\end{equation}}
\newcommand{\bw}{\begin{widetext}}
\newcommand{\ew}{\end{widetext}}
\newcommand{\bi}{\begin{itemize}}
\newcommand{\ei}{\end{itemize}}
\newcommand{\ud}{\mathrm{d}}

\newcommand{\LCm}{{\scriptscriptstyle -}} \newcommand{\LCp}{{\scriptscriptstyle +}}
\newcommand{\LCpm}{{\scriptscriptstyle \pm}}
\newcommand{\LCmp}{{\scriptscriptstyle \mp}}

\newcommand{\LCperp}{{\scriptscriptstyle \perp}}

\newcommand{\LCpara}{{\scriptscriptstyle \parallel}}

\usepackage[T1]{fontenc} \usepackage[latin1]{inputenc}

\begin{document}

\title{Quantum radiation reaction spectrum of electrons in plane waves}

\author{Greger Torgrimsson}
\email{greger.torgrimsson@umu.se}
\affiliation{Department of Physics, Ume{\aa} University, SE-901 87 Ume{\aa}, Sweden}

\begin{abstract}

In previous works we derived equations for the average momentum of high-energy electrons experiencing quantum radiation reaction (RR) in strong electromagnetic plane-wave background fields. In this paper we derive similar equations for the momentum spectrum. We formulate the equations in terms of the cumulative function and study the relation between the equations for the spectrum and the equations for the moments, analyze the structure of the low-energy expansions, and finally explain how our formulation is essentially in terms of a ``Green's function'' which allows us to study the dynamics without choosing a specific initial wave packet (or particle-bunch distribution).  

\end{abstract}
\maketitle

\section{Introduction}

Electrons and positrons can experience significant quantum RR in high-intensity laser fields. Standard methods to study this include particle-in-cell (PIC) codes, kinetic equations~\cite{Shen1972,Sokolov:2010am,Neitz:2013qba,Neitz:2014hla,NielPRE2018,Elkina:2010up,Seipt:2023bcw}, or by rescaling the RR term in the Landau-Lifshitz (LL) equation with the so-called Gaunt factor. For reviews of RR and other phenomena in strong field QED see \cite{DiPiazza:2011tq,Gonoskov:2021hwf,Fedotov:2022ely}.
While these methods are standard, their precision has not been fully checked against experiments, because there have so far only been very few experiments, \cite{Poder:2017dpw,Cole:2017zca,Wistisen:2017pgr}, which actually suggested there might be some discrepancies, at least in the considered parameter regime.

Historically, spin and polarization have usually been neglected. However, since~\cite{DelSorbo:2017fod,DelSorbo2,Seipt:2019ddd,Li:2018fcz} it has been realized that, not only is this often not negligible, but lasers can be useful for generating polarized particle beams. 
Spin and polarization have been included in PIC codes using Stokes vectors in~\cite{Li:2018fcz,Li:2019oxr,Li:2020bwo,Tang:2021azl,Li:2022wqn}, and Stokes vectors and Mueller matrices have recently been used in~\cite{Seipt:2023bcw} to derive kinetic equations.

In this paper we will derive RR directly from quantum field theory. We started this in~\cite{Torgrimsson:2021wcj,Torgrimsson:2021zob}, where we derived new equations for the average momentum and for the spin transition probability. Here we will derive similar equations for the momentum spectrum. 
The starting point is, as usual in QFT, an expansion of the relevant probabilities in a power series in $\alpha$. Since we are interested in strong fields, this is done in the Furry picture, where we keep the exact dependence on the background field. Since the background field is always accompanied by a factor of the charge, but since this factor is not part of the expansion in $\alpha$, we rescale the background field as $eF_{\mu\nu}\to F_{\mu\nu}$. We are interested in regimes where higher orders in $\alpha$ are important, which means we have to resum the $\alpha$ expansion. There have been several other types of resummations of the Furry-picture expansion for other regimes, quantities and processes in recent papers~\cite{Heinzl:2021mji,Ekman:2021eqc,Karbstein:2019wmj,Mironov:2020gbi,Mironov:2022jbg,Edwards:2020npu,Podszus:2021lms,Podszus:2022jia}.    

Whether we are calculating the average momentum, spin transition, or the spectrum, in this approach we are always calculating some inclusive probability. Here we have the inclusive probability that an electron ends up with a certain momentum and spin given some initial momentum and spin, summed over the probabilities that this happens after emitting 0, 1, 2... photons. The probability that $n$ photons are emitted is itself given by an infinite sum over loops. Clearly, one cannot calculate multi-photon emissions or higher-order loops exactly, even after simplifying by assuming the background field is a plane wave (for which the solution to the Dirac equation is particularly simple). In~\cite{Dinu:2018efz,Dinu:2019pau,Torgrimsson:2020gws} we showed how higher-orders-in-$\alpha$ processes can be approximated by incoherent products of ``strong-field-QED Mueller matrices''. In~\cite{Torgrimsson:2021wcj,Torgrimsson:2021zob} we showed how to sum the individual probabilities and how to actually evaluate the sum. The results are recursive and integrodifferential matrix equations. To summarize some of the differences between ours and other approaches, we do not introduce rates, our equations are formulated in terms of a single electron rather than an electron bunch described by a classical particle distribution, and we are not restricted to the locally-constant-field (LCF) regime.     

As a motivation for considering plane waves, note that an electron with sufficiently high energy will effectively see a more general background field as if it were a plane wave, which follows from a short Lorentz boost argument. There are of course exceptions to this. For example, if, on the trajectory of the electron, the only nonzero component of the background field is an electric field parallel to the electron momentum, see e.g.~\cite{Brodin:2022dkd,DegliEsposti:2023qqu}. But in general one can expect a plane wave to be a good approximation of a general field.

In the literature it is common to simplify further by treating the plane wave in a LCF approximation. This additional step requires that\footnote{We use units with the electron mass $m_e=1$ in addition to $c=\hbar=1$.} $a_0=E/\omega$ is sufficiently large, where $E$ is the maximum field strength and $\omega$ a characteristic frequency scale. In Sec.~\ref{Derivation} and~\ref{Finite wave packets} we derive general equations that are valid beyond the LCF regime, see in particular e.g.~\eqref{integroDiffx}, \eqref{MnEq}, \eqref{MfFromM} and~\eqref{integroSf}. The general Mueller matrices can also be expressed compactly in terms of known special functions if, instead of assuming large $a_0$, one assumes that the field has circular polarization and is sufficiently long, which allows one to use a locally monochromatic field approximation, see~\cite{Torgrimsson:2020gws}. For different ways of treating the field as locally monochromatic, see~\cite{Heinzl:2020ynb,Blackburn:2023mlo}. 

Another important parameter is $\chi=\sqrt{-(F^{\mu\nu}p_\nu)^2}$, where $p_\mu$ is the electron momentum. Quantum effects can be neglected for sufficiently small $\chi$. Here we are interested in values of $\chi$ that are large enough for significant quantum effects in RR, but small enough so that we can neglect pair production. To lowest order the probability of trident pair production scales as~\cite{Ritus:1972nf,Baier} $\mathbb{P}(e^{\LCm}\to e^{\LCm} e^{\LCm} e^{\LCp})\propto\exp(-16/[3\chi])$. In~\cite{Torgrimsson:2021wcj,Torgrimsson:2021zob} we studied the $\chi\ll1$ expansions of the average momentum and the spin transition probability, i.e. the first two moments of the spectrum, and showed that they are asymptotic and can be resummed with the Borel-Pad\'e method. In Sec.~\ref{Constant field} we show how to obtain $\chi\ll1$ expansions of the spectrum. We find that the expansion parameter is $\sqrt{\chi}$, so there is significant room for quantum effects in RR while pair production is still negligible.

\section{Derivation}\label{Derivation}

Lightfront coordinates are defined as
\be
v^\LCpm=2v_\LCmp=v^0\pm x^3
\ee
and $v_\LCperp=\{v_1,v_2\}$, so that the background field is given by $a_\LCperp(\phi)$ and $a_\LCpm=0$, where $\phi=kx=\omega x^\LCp$. We consider general pulse shape and polarization. We are interested in the dependence on the lightfront longitudinal momentum, $kP$. Rather than considering the spectrum directly, we consider instead a partially integrated spectrum or a cumulative distribution function, which we define as the probability that an electron which initially has longitudinal momentum $kp$ emerges, after interaction with the background field, with momentum
\be\label{xDefinition}
kp'>(1-x)kp \;,
\ee
where $0<x<1$. We also consider general spin transition. If we sum over the final spin and set $x=1$ then we obtain a probability $\mathbb{P}=1$ as we should. $x=0$ gives the probability of electrons that have lost no or almost no longitudinal momentum. Differentiating the final result with respect to $x$ gives the spectrum.
As our starting point is the Furry-picture expansion in $\alpha$, we initially have the probability as a Taylor expansion in $\alpha$,
\be
\mathbb{P}=\sum_{n=0}^\infty\mathbb{P}^{(n)} \;,
\ee
where each $\mathbb{P}^{(n)}=\mathcal{O}(\alpha^n)$ is a nontrivial function of the field strength\footnote{Recall that we have absorbed a factor of $e$ as $eE\to E$.}. We can write each term as a multiplication of the (4D) Stokes vectors, ${\bf N}_0$ and ${\bf N}_f$, for the initial and final spin and a ($4\times4$) Mueller matrix,
\be
\mathbb{P}^{(n)}=\frac{1}{2}{\bf N}_0\cdot{\bf M}^{(n)}\cdot{\bf N}_f \;.
\ee
Note that we first calculate ${\bf M}^{n}$ or
\be
{\bf M}=\sum_{n=0}^\infty{\bf M}^{n} \;,
\ee
so we do not need to choose any specific initial or final spin until the very end of the calculation, where we simply have to project the result for ${\bf M}$ with the Stokes vectors,
\be
\mathbb{P}=\frac{1}{2}{\bf N}_0\cdot{\bf M}\cdot{\bf N}_f \;.
\ee

The initial state is described by a wave packet as
\be\label{inPacket}
|\text{in}\rangle=\int\ud\tilde{p}_0f(p_0)b^\dagger(p_0,s)|0\rangle \;,
\ee
where
\be
\ud\tilde{p}=\frac{\theta(p_\LCm)\ud p_\LCm\ud^2 p_\LCperp}{(2\pi)^32p_\LCm}
\ee
is the usual Lorentz-invariant integration measure, written here in lightfront coordinates, and
\be
1=\langle\text{in}|\text{in}\rangle=\int\ud\tilde{p}|f|^2 \;.
\ee
We consider first a wave packet which is sharply peaked, and then in Sec.~\ref{Finite wave packets} we show that the results for a sharply peaked wave packet essentially give what can be thought of as a Green's function, from which one can afterwards obtain the results for an arbitrary, wide wave packet.
A general spin can be written as
\be\label{rholambda}
b^\dagger|0\rangle=\cos\left(\frac{\rho}{2}\right)b_\uparrow^\dagger|0\rangle+\sin\left(\frac{\rho}{2}\right)e^{i\lambda}b_\downarrow^\dagger|0\rangle \;,
\ee
where $\rho$ and $\lambda$ are two real constants. To zeroth order we hence find
\be\label{Pzeroth}
\begin{split}
\mathbb{P}^{(0)}&=\int\ud\tilde{p}'\theta(kp'-[1-x]kp)|\langle0|b(p',\rho_1,\lambda_1)|\text{in}\rangle|^2 \\
&=\int\ud\tilde{p}'\theta(kp'-[1-x]kp)|f|^2\frac{1}{2}{\bf N}_0\cdot{\bf N}_1 \\
&\to\frac{1}{2}{\bf N}_0\cdot{\bf N}_1 \;,
\end{split}
\ee
where the Stokes vectors are related to the angles in~\eqref{rholambda} as
\be
{\bf N}=\{1,\cos\lambda\sin\rho,\sin\lambda\sin\rho,\cos\rho\} \;.
\ee

Here we have assumed that the wave packet is sufficiently narrow compared the values of $x$ we consider. 
If we were to consider $x$ too close to $0$, then the step function would essentially be $\theta(kp'-kp)$ and we would find
\be
\int\ud\tilde{p}'\theta(kp'-kp)|f(p')|^2\sim\frac{1}{2}
\ee
as only about one half of the peak of $f$ would be integrated over. For example, if we were to consider a Gaussian wave packet proportional to
\be
f(p')\propto\exp\left[-\frac{(kp'-kp)^2}{\lambda^2}\right] \;,
\ee
then for sufficiently small $\lambda$ we would find 
\be
\mathbb{P}^{(0)}(\lambda)\propto\frac{1}{2}\left(1+\text{erf}\left[\frac{xkp}{\lambda}\right]\right) \;.
\ee
This is nonzero but exponentially suppressed for $x<0$. A nonzero $\lambda$ gives a regularized step function for the cumulative distribution, and a regularized delta function for the spectrum. 
However, until Sec.~\ref{Finite wave packets}, we will assume that the wave packet is sufficiently narrow so that we can always approximate
\be
\int\ud\tilde{p}'|f(p')|^2h(p')\approx h(p) \;.
\ee
Thus, from~\eqref{Pzeroth} we have
\be
{\bf M}^{(0)}={\bf 1} \;.
\ee

At $\mathcal{O}(\alpha)$ we have the loop correction to scattering without emission and emission of one photon without loops, which can be written as ${\bf M}^{(1)}(b_0,x,-\infty)$, where~\cite{Dinu:2019pau,Torgrimsson:2020gws}
\be\label{M1fromMLMC}
{\bf M}^{(1)}(b_0,x,\sigma)=\int_\sigma^\infty\ud\sigma'\int_0^1\ud q\left[{\bf M}_L+\theta(x-q){\bf M}_C\right]
\ee
where $\sigma=(\phi_2+\phi_1)/2$ is the average lightfront time, with $\phi_2$ being the lightfront-time variable for the amplitude $M$ and $\phi_1$ for the complex conjugate $\bar{M}$, $q=kl/kp=kl/b_0$ is the ratio of the longitudinal momentum of the photon and the initial electron, and ${\bf M}_C$ and ${\bf M}_L$ are the Mueller matrices for photon emission and the electron self-energy loop. The reason for introducing the lower integration limit for the lightfront-time integral, $\theta(\sigma'-\sigma)$, will be explained below. The step function $\theta(x-q)$ is due to the restriction in~\eqref{xDefinition}, which to this order and for this term can be rearranged into an upper cut-off for the photon momentum, $q<x$. There is no such step function for ${\bf M}_L$ because that $q$ integral corresponds to a photon that is emitted and reabsorbed  and therefore does not change the final electron momentum.

For an arbitrary field, ${\bf M}_C$ is given by Eq.~(24), (25), (26), (27) and (29) in~\cite{Dinu:2019pau}, and ${\bf M}_L$ by Eq.~(67) to (71) in~\cite{Torgrimsson:2020gws}, which we write here as\footnote{When comparing the overall factor of $2$ in Eq.~(24) in~\cite{Dinu:2019pau}, note that here we get one extra factor of $2$ when summing over the photon polarization, and one factor of $1/2$ has been factored out, writing $\mathbb{P}^{(1)}=(1/2){\bf N}_0\cdot{\bf M}^{(1)}\cdot{\bf N}_f$.}
\be\label{thetaInt}
{\bf M}_{C,L}(b_0,q,\sigma)=\frac{i\alpha}{2\pi b_0}\int\frac{\ud\theta}{\theta}\exp\left\{\frac{ir}{2b_0}\theta M_{21}^2\right\}{\bf R}_{C,L} \;,
\ee 
where $\theta=\phi_2-\phi_1$, the integration contour goes above the pole at $\theta=0$, $r=(1/s)-1$, $s=1-q$ is the ratio of the longitudinal momentum of the electron after and before emitting the photon, $b_0=kp$, and $M_{21}^2$ is an effective mass
\be
M_{ij}^2=\langle\pi\rangle_{ij}^2=1+\langle{\bf a}^2\rangle_{ij}-\langle{\bf a}\rangle_{ij}^2 \;,
\ee
where
\be
\langle F\rangle_{ij}=\frac{1}{\theta_{ij}}\int_{\phi_j}^{\phi_i}\ud\phi\, F(\phi) \;.
\ee
For photon emission, the $4\times4$ matrix is given by
\be
{\bf R}_C=\begin{pmatrix}\langle\mathbb{R}^C\rangle & {\bf R}_1^C\\
{\bf R}_0^C & {\bf R}_{01}^C \end{pmatrix} \;,
\ee
where
\be
\langle\mathbb{R}^C\rangle=\frac{\kappa}{2}\left[\frac{2ib_0}{r\theta}+1+D_1\right]-1 \;,
\ee
\be
{\bf R}_0^C=q\left\{{\bf 1}+\left[1+\frac{1}{s}\right]\hat{\bf k}\,{\bf X}\right\}\!\cdot\!{\bf V} \;,
\ee
\be
{\bf R}_1^C=\frac{q}{s}\left\{{\bf 1}+\left[1+s\right]\hat{\bf k}\,{\bf X}\right\}\!\cdot\!{\bf V}
\ee
and
\be
\begin{split}
{\bf R}_{01}^C=&\frac{q}{s}\left\{\hat{\bf k}\,{\bf X}-s{\bf X}\,\hat{\bf k}-\frac{q}{2}\hat{\bf k}\,\hat{\bf k}\right\}\\
&+\left[\frac{2ib_0}{r\theta}+D_1\right]\left[{\bf 1}_2+\frac{\kappa}{2}\hat{\bf k}\,\hat{\bf k}\right] \;,
\end{split}
\ee
where $\kappa=(1/s)+s$, $D={\bf w}_1\cdot{\bf w}_2$, ${\bf 1}_2={\bf 1}-\hat{\bf k}\hat{\bf k}$, 
\be
{\bf X}=\frac{1}{2}({\bf w}_2+{\bf w}_1)
\qquad
{\bf V}=\frac{1}{2}{\bm\sigma}_2\!\cdot\!({\bf w}_2-{\bf w}_1) \;,
\ee
\be
{\bm\sigma}_2=\begin{pmatrix}0&-i\\ i&0\end{pmatrix}
\ee
and
\be
{\bf w}_1={\bf a}(\phi_1)-\langle{\bf a}\rangle_{21}
\qquad
{\bf w}_2={\bf a}(\phi_2)-\langle{\bf a}\rangle_{21} \;.
\ee
Note that no special notation has been used for outer products, so, for example, $(\hat{\bf k}\,{\bf X}\!\cdot\!{\bf V})_j=\hat{\bf k}_j({\bf X}\!\cdot\!{\bf V})$.
For the loop we have
\be
{\bf R}_L=\begin{pmatrix}\langle\mathbb{R}^L\rangle & {\bf R}_1^L\\
{\bf R}_0^L & {\bf R}_{01}^L \end{pmatrix}=\begin{pmatrix}-\langle\mathbb{R}^C\rangle &-{\bf R}_0^C\\-{\bf R}_0^C &-\langle\mathbb{R}^C\rangle{\bf 1}+{\bf R}_{01}^{\rm rot} \end{pmatrix} \;,
\ee
so some of the elements are identical to ${\bf R}_C$, while
\be
{\bf R}_{01}^{\rm rot}=\text{sign}(\theta)\left[\frac{q}{2}({\bf Y}\hat{\bf k}-\hat{\bf k}{\bf Y})-q\left[1+\frac{1}{s}\right]({\bf X}\!\cdot\!{\bf V}){\bm\sigma}_2\right] \;,
\ee
where ${\bf Y}={\bf w}_2-{\bf w}_1$, gives spin rotation.
Note that 
\be
({\bf M}_L+{\bf M}_C)\!\cdot\!{\bf e}_0=0 \;,
\ee
where
\be
{\bf e}_0=\{1,0,0,0\} \;,
\ee
so if we sum over the final spin state and if we integrate over all momenta, i.e. $x=1$, then $\mathbb{P}^{(1)}=0$ for any initial spin state. We also have $\mathbb{P}^{(n>1)}=0$, so $\mathbb{P}=\mathbb{P}^{(1)}={\bf N}_0\!\cdot\!{\bf e}_0=1$, which is what it has to be due to unitarity.

To $\mathcal{O}(\alpha^2)$ we have
\be\label{firstM2eq}
\begin{split}
{\bf M}^{(2)}&=\int_\sigma^\infty\ud\sigma_1\int_0^1\ud q_1\int_{\sigma_1}^\infty\ud\sigma_2\int_0^1\ud q_2\\
&\times\bigg\{{\bf M}_L(b_0,q_1,\sigma_1)\!\cdot\!{\bf M}_L(b_0,q_2,\sigma_2)\\
&+{\bf M}_L(b_0,q_1,\sigma_1)\!\cdot\!\theta(x-q_2){\bf M}_C(b_0,q_2,\sigma_2)\\
&+\theta(x-q_1){\bf M}_C(b_0,q_1,\sigma_1)\!\cdot\!{\bf M}_L([1-q_1]b_0,q_2,\sigma_2)\\
&+\theta(x-q_1){\bf M}_C(b_0,q_1,\sigma_1)\\
&\hspace{.25cm}\cdot\theta\left(\frac{x-q_1}{1-q_1}-q_2\right){\bf M}_C([1-q_1]b_0,q_2,\sigma_2)\bigg\} \;,
\end{split}
\ee
where, in the last term, we have used
\be\label{stepFun2}
\theta(kp'-[1-x]kp)=\theta(x-q_1)\theta\left(\frac{x-q_1}{1-q_1}-q_2\right) \;,
\ee
where (in this term) $kp'=kp-kl_1-kl_2=kp_1-kl_2$, $q_1=kl_1/kp$, $q_2=kl_2/kp_1$, and $kp_1/kp=1-q_1$. Note that $q_j$ is the longitudinal momentum of the photon emitted (or emitted and reabsorbed) at step $n$ divided by the longitudinal momentum of the electron immediately before this step. Writing the step function as in~\eqref{stepFun2} allows us to obtain ${\bf M}^{(2)}$ from ${\bf M}^{(1)}$ by prepending ${\bf M}_L$ and ${\bf M}_C$ as
\be\label{secondM2eq}
\begin{split}
{\bf M}^{(2)}(b_0,x,\sigma)&=\int_\sigma^\infty\ud\sigma_1\int_0^1\ud q_1\\
&\times\bigg\{{\bf M}_L(b_0,q_1,\sigma_1)\!\cdot\!{\bf M}^{(1)}(b_0,x,\sigma_1)\\
&+\theta(x-q_1){\bf M}_C(b_0,q_1,\sigma_1)\\
&\!\cdot\!{\bf M}^{(1)}\left([1-q_1]b_0,\frac{x-q_1}{1-q_1},\sigma_1\right)\bigg\} \;.
\end{split}
\ee
Note that if we replace ${\bf M}^{(2)}\to{\bf M}^{(1)}$ and ${\bf M}^{(1)}\to{\bf M}^{(0)}={\bf 1}$ then we have~\eqref{M1fromMLMC}.
Higher orders can be obtained recursively\footnote{A recursive formula relating the probability $P_n$ to emit $n$ photons to $P_{n-1}$ has also been derived in~\cite{Tamburini:2019tzo}.} from
\be
\begin{split}
&{\bf M}^{(n)}(b_0,x,\sigma)=\int_\sigma^\infty\ud\sigma'\int_0^1\ud q\\
&\times\bigg\{{\bf M}^L(b_0,q,\sigma')\!\cdot\!{\bf M}^{(n-1)}(b_0,x,\sigma')+\\
&\theta(x-q){\bf M}^C(b_0,q,\sigma')\!\cdot\!{\bf M}^{(n-1)}\left[(1-q)b_0,\frac{x-q}{1-q},\sigma'\right]\bigg\} \;.
\end{split}
\ee
Summing over $n\geq1$ and differentiating with respect to $\sigma$ gives
\be\label{integroDiffx}
\begin{split}
\frac{\partial{\bf M}}{\partial\sigma}=&-\int_0^1\ud q\bigg\{{\bf M}^L\!\cdot\!{\bf M}(b_0,x)\\
&+\theta(x-q){\bf M}^C\!\cdot\!{\bf M}\left([1-q]b_0,\frac{x-q}{1-q}\right)\bigg\} \;,
\end{split}
\ee
where we have suppressed the lightfront time argument $\sigma$ as it is now the same in all terms. We have an ``initial'' condition at $\sigma\to+\infty$ rather than $\sigma\to-\infty$,
\be
{\bf M}(\sigma\to+\infty)={\bf M}^{(0)}={\bf 1} \;,
\ee
and then we integrate backwards in lightfront time, where the final result is given by ${\bf M}(\sigma\to-\infty)$.

An alternative form that might be illuminating is obtained by using $b':=(1-x)b_0$ instead of $x$ for the final momentum, and by changing integration variable from $q$ to $b=(1-q)b_0$,
\be\label{Mintegrobprime}
\begin{split}
\frac{\partial}{\partial\sigma}{\bf M}(b_0,b')=&-\int_0^{b_0}\frac{\ud b}{b_0}\bigg\{{\bf M}^L\!\cdot\!{\bf M}(b_0,b')\\
&+\theta(b-b'){\bf M}^C\!\cdot\!{\bf M}(b,b')\bigg\} \;.
\end{split}
\ee

As a check, for $x=1$, which means no restriction on the final momentum, we have
\be\label{eqFor1}
\begin{split}
&\frac{\partial}{\partial\sigma}{\bf M}(b_0,1)=\\
&-\int_0^1\ud q\bigg\{{\bf M}^L\!\cdot\!{\bf M}(b_0,1)+{\bf M}^C\!\cdot\!{\bf M}\left([1-q]b_0,1\right)\bigg\} \;,
\end{split}
\ee
which is the same equation as in~\cite{Torgrimsson:2021wcj,Torgrimsson:2021zob}.

For $x=0$, which means observing only electrons that have lost no or very little longitudinal momentum, the photon-emission term becomes negligible due to $\theta(x-q)$, and we have
\be
\frac{\partial}{\partial\sigma}{\bf M}(x=0)=-\int_0^1\ud q\,{\bf M}_L\cdot{\bf M}(x=0) \;.
\ee
The solution is given by a lightfront-time-ordered exponential,
\be\label{timeOrderedExpx0}
\begin{split}
{\bf M}(x=0)&={\bf 1}+\sum_{n=1}^\infty\int_\sigma^\infty\ud\sigma_1\int_{\sigma_1}^\infty\ud\sigma_2...\int_{\sigma_{n-1}}^\infty\ud\sigma_n\\
&\times\left(\int_0^1\!\ud q\,{\bf M}_L(\sigma_1)\right)...\left(\int_0^1\!\ud q\,{\bf M}_L(\sigma_n)\right)\\
&=\bar{T}\exp\left(\int_\sigma^\infty\!\ud\sigma'\int_0^1\!\ud q\,{\bf M}_L(\sigma')\right) \;,
\end{split}
\ee
where $\bar{T}$ means anti-lightfront-time ordering. This agrees with the result in~\cite{Torgrimsson:2020gws}. As shown in~\cite{Torgrimsson:2020gws}, for a field with linear or circular polarization, one can write the result in an explicit form without a time-ordered exponential of a matrix. Thus, in general, the new equation~\eqref{integroDiffx} interpolates between the result in~\cite{Torgrimsson:2020gws} for $x=0$ and the one in~\cite{Torgrimsson:2021wcj,Torgrimsson:2021zob} for $x=1$.

\subsection{Moments}

The spectrum is given by 
\be\label{Sdefx}
{\bf S}:=\frac{\partial{\bf M}}{\partial x} \;,
\ee
from which we can obtain the moments 
\be
\langle (kP)^n\rangle=\frac{1}{2}{\bf N}_0\cdot\tilde{\bf M}\cdot{\bf N}_f 
\ee
as
\be\label{momentsdMdxOneMinusx}
\tilde{\bf M}(n,b_0)=b_0^n\int_{-\infty}^1\ud x\frac{\partial{\bf M}}{\partial x}(1-x)^n \;.
\ee
The lower integration limit is not $x=0$ because then we would miss the essentially delta-function-like peak at $x=0$. The width of the wave packet determines the width of this peak. For a sharply peaked wave packet the probability for $x<0$ is very small but it is nonzero. We can avoid the $x<0$ region by making a partial integration,
\be\label{MnFromIntM}
\begin{split}
\tilde{\bf M}(b_0)&=nb_0^n\int_{-\infty}^1\ud x(1-x)^{n-1}{\bf M}\\
&\to nb_0^n\int_0^1\ud x(1-x)^{n-1}{\bf M} \;,
\end{split}
\ee
where in the second step we have neglected the $x<0$ part since now the integrand has no delta-function-like peak and is negligibly small for $x<0$ (because we have assumed a sharply peaked wave packet). We can integrate~\eqref{integroDiffx} over $x$ before solving it, i.e. we can express
\be
nb_0^n\int_0^1\ud x(1-x)^{n-1}\eqref{integroDiffx} 
\ee
entirely in terms of $\tilde{\bf M}$.
This is trivial for the first two terms, while for third we change variable from $x$ to $x'=(x-q)/(1-q)$. We find
\be\label{MnEq}
\frac{\partial}{\partial\sigma}\tilde{\bf M}(b_0)=
-\int_0^1\!\ud q\bigg\{{\bf M}^L\!\cdot\!\tilde{\bf M}(b_0)+{\bf M}^C\!\cdot\!\tilde{\bf M}\left([1-q]b_0\right)\bigg\} \;.
\ee
We thus have the same equation for all $n$. $n$ only appears in the ``initial'' condition
\be\label{initialMn}
\tilde{\bf M}(\sigma\to+\infty,n,b_0)=b_0^n{\bf 1} \;.
\ee
Note that~\eqref{MnEq} allows us to obtain each moment separately. In other words, if we want, say, the second moment then we do not need to consider any other moments and we do not need to obtain the spectrum. 
For $n=0$ we have $\tilde{\bf M}(n=0)={\bf M}(x=1)$ and so~\eqref{MnEq} is the same equation as~\eqref{eqFor1}. $n=1$ gives the expectation value $\langle kP\rangle$ which we studied in~\cite{Torgrimsson:2021wcj}. \eqref{MnEq} together with~\eqref{initialMn} are what one should expect by a generalization of the derivation for $n=1$ in~\cite{Torgrimsson:2021wcj}, but this indirect derivation of~\eqref{MnEq} from~\eqref{integroDiffx} serves as a check of~\eqref{MnFromIntM}. 

Apart from checking that these two sets of equations are consistent, this also allows us to check the numerical results obtained from them. After having obtained a numerical solution of ${\bf M}(x)$ (for $x>0$) we can check the result by integrating it as in~\eqref{MnFromIntM} for the first couple of moments and comparing with the moments obtained by instead using~\eqref{MnEq}, which is much faster to solve since $\tilde{\bf M}$ only has two integration variables, $\sigma$ and $b_0$, while ${\bf M}$ also has $x$.  

Thus, all the moments obtained from ${\bf M}(b_0,x)$ by solving the new equation~\eqref{integroDiffx} and using \eqref{MnFromIntM} agree with the moments obtained using the approach in~\cite{Torgrimsson:2021wcj}. Two different distributions can in principle have the same moments\footnote{We would be dealing with Stieltjes (Hausdorff) moment problem for $-\infty<x<1$ ($0<x<1$).}. However, in the above derivation we did not need to assume that $n$ is an integer. If we let $n$ be a continuous variable then we essentially have the Mellin transform of the spectrum. The inverse would be given by an integral over $n$ in the complex plane. Or we could replace $(1-x)^nb_0^n$ in~\eqref{momentsdMdxOneMinusx} with some arbitrary function $f([1-x]b_0)$. If $f(0)$ is nonzero then we can simply deal with that constant separately, so we can assume without loss of generality that $f(0)=0$. Now the resulting $\tilde{\bf M}$ is determined by the same equation as the moments, i.e.~\eqref{MnEq}, but with ``initial'' condition $\tilde{M}(\sigma\to+\infty,b_0)=f(b_0){\bf1}$. Thus, the result is again what we would find if we instead started with the approach in~\cite{Torgrimsson:2021wcj}.

\subsection{Locally constant field approximation}

For sufficiently large $a_0$ we can use a LCF approximation. Here the $\theta$ integrals in~\eqref{thetaInt} can be performed in terms of Airy functions, $\text{Ai}$, $\text{Ai}'$ and
\be
\text{Ai}_1(\xi)=\int_\xi^\infty\ud t\,\text{Ai}(t) \;,
\ee
and the Scorer function $\text{Gi}$, 
\be
\frac{\text{Ai}(\xi)+i\text{Gi}(\xi)}{\sqrt{\xi}}=\int_0^\infty\frac{\ud\tau}{\pi}\exp\left\{i\xi^{3/2}\left(\tau+\frac{\tau^3}{3}\right)\right\} \;.
\ee
From~\cite{Torgrimsson:2020gws} we have
\be
\begin{split}
{\bf M}_C&=\frac{\alpha}{b_0}\bigg\{-\left(\text{Ai}_1(\xi)+\kappa\frac{\text{Ai}'(\xi)}{\xi}\right){\bf e}_0{\bf e}_0\\
&+\frac{q}{s}\frac{\text{Ai}(\xi)}{\sqrt{\xi}}{\bf e}_0\hat{\bf B} +
q\frac{\text{Ai}(\xi)}{\sqrt{\xi}}\hat{\bf B}{\bf e}_0\\
&-\text{Ai}_1(\xi)({\bf 1}_\LCperp+[\kappa-1]{\bf 1}_\LCpara)-\frac{\text{Ai}'(\xi)}{\xi}(2{\bf 1}_\LCperp+\kappa{\bf 1}_\LCpara)\bigg\} \;,
\end{split}
\ee
and
\be
\begin{split}
{\bf M}_L&=\frac{\alpha}{b_0}\bigg\{\left(\text{Ai}_1(\xi)+\kappa\frac{\text{Ai}'(\xi)}{\xi}\right){\bf 1}_4\\
&-q\frac{\text{Ai}(\xi)}{\sqrt{\xi}}({\bf e}_0\hat{\bf B}+\hat{\bf B}{\bf e}_0)
+q\frac{\text{Gi}(\xi)}{\sqrt{\xi}}(\hat{\bf k}\hat{\bf E}-\hat{\bf E}\hat{\bf k})\bigg\} \;,
\end{split}
\ee
where $\xi=(r/\chi(\sigma))^{2/3}$, $\chi(\sigma)=|{\bf a}'(\phi)|b_0$, 
\be
\begin{split}
{\bf e}_0&=\{1,0,0,0\} 
\qquad
{\bf e}_1=\{0,1,0,0\} \\
{\bf e}_2&=\{0,0,1,0\} 
\qquad
{\bf e}_3=\{0,0,0,1\} \;,
\end{split} 
\ee
the local directions of the electric and magnetic are defined initially as 3D vectors as
\be
\hat{\bf E}(\sigma)=\frac{{\bf a}'(\sigma)}{|{\bf a}'(\sigma)|}
\qquad
\hat{\bf B}(\sigma)=\hat{\bf E}(\sigma)\times\hat{\bf k} \;,
\ee
but made 4D trivially as
\be
\hat{\bf E}(\sigma)=E_1(\sigma){\bf e}_1+E_2(\sigma){\bf e}_2
\ee
and similar for $\hat{\bf B}(\sigma)$, $\hat{\bf k}={\bf e}_3$, ${\bf 1}_4=\sum_{j=1}^4{\bf e}_j{\bf e}_j$ and
\be
{\bf 1}_\LCperp=\hat{\bf E}\hat{\bf E}+\hat{\bf B}\hat{\bf B}={\bf e}_1{\bf e}_1+{\bf e}_2{\bf e}_2
\qquad
{\bf 1}_\LCpara={\bf e}_3{\bf e}_3 \;.
\ee

The ${\bf e}_0\hat{\bf B}$ and $\hat{\bf B}{\bf e}_0$ terms lead to induced polarization along the magnetic field direction, as a Sokolov-Ternov effect, while the $\hat{\bf k}\hat{\bf E}-\hat{\bf E}\hat{\bf k}$ term gives spin rotation, including the effect of the anomalous magnetic moment (cf.~\cite{Ilderton:2020gno,BaierSokolovTernov}). For a rotating field, we have in general $\hat{\bf B}(\sigma_1)\cdot\hat{\bf E}(\sigma_2)\ne0$, so all $4$ components couple, and an induced polarization along the local direction of the magnetic field will later be along the electric field direction, which undergoes spin rotation. However, for a linearly polarized field, the ${\bf e}_3$ and $\hat{\bf E}$ components decouple from the ${\bf e}_0$ and $\hat{\bf B}$ components.
When we for such fields consider initial or final Stokes vectors with ${\bf e}_3\cdot{\bf N}_{0,f}=\hat{\bf E}_3\cdot{\bf N}_{0,f}=0$, then we drop the irrelevant components and use 2D Stokes vectors and $2\times2$ Mueller matrices.

\section{Constant field}\label{Constant field}

\begin{figure}
\includegraphics[width=\linewidth]{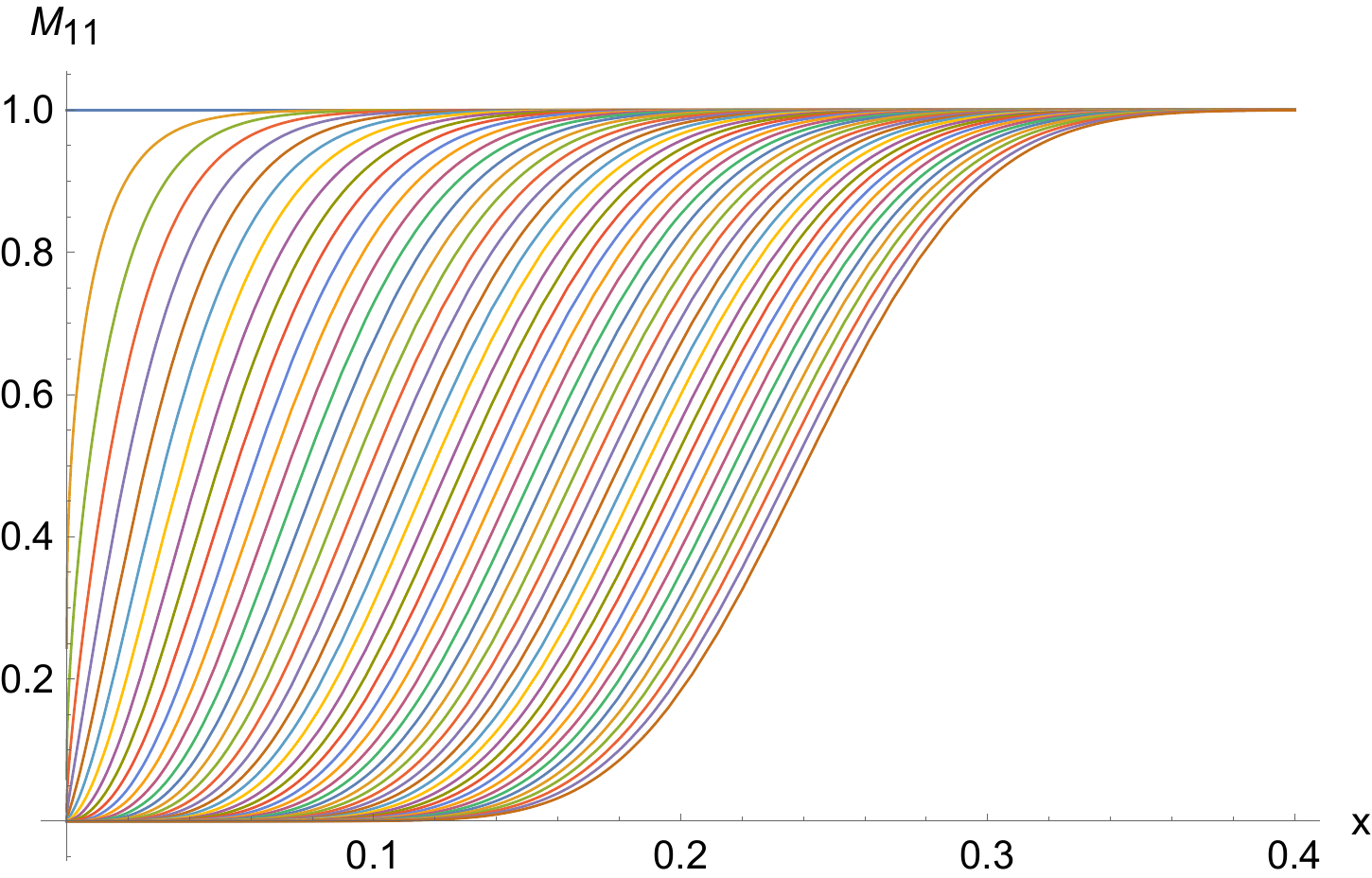}
\includegraphics[width=\linewidth]{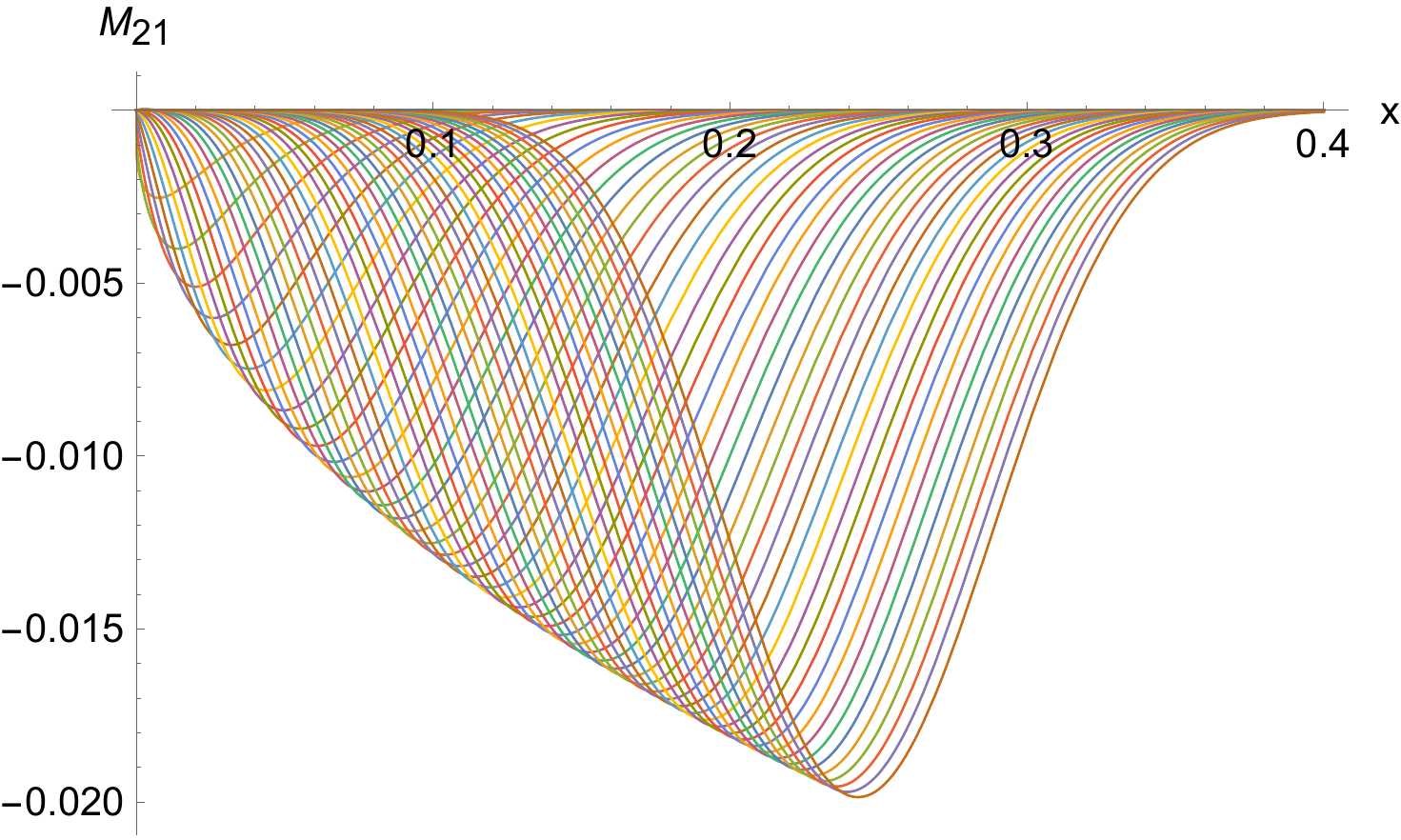}
\caption{Cumulative momentum distribution, $M_{11}=\{1,0\}\cdot{\bf M}\cdot\{1,0\}$ and $M_{21}=\{0,1\}\cdot{\bf M}\cdot\{1,0\}$ for $\chi=0.01$ and for $T=0,1,2\dots50$.}
\label{Mchi0p01fig}
\end{figure}

For a constant field we have
\be
\int_\sigma^\infty\ud\sigma_1\int_{\sigma_1}^\infty\ud\sigma_2...\int_{\sigma_{n-1}}^\infty\ud\sigma_n=\frac{\Delta\phi^n}{n!} \;,
\ee
so we have an effective expansion parameter
\be
T=\alpha a_0\Delta\phi \;.
\ee
We separate $T^n$ from ${\bf M}^{(n)}$, changing notation slightly so that
\be
{\bf M}=\sum_{n=0}^\infty T^n{\bf M}^{(n)} \;.
\ee
We also separate a factor of $\alpha a_0$ from ${\bf M}_{L,C}$, and then the recursive formula simplifies
\be
\begin{split}
&{\bf M}^{(n)}(b_0,x)=\int_0^1\frac{\ud q}{n}\bigg\{{\bf M}^L(b_0,q)\!\cdot\!{\bf M}^{(n-1)}(b_0,x)\\
&+\theta(x-q){\bf M}^C(b_0,q)\!\cdot\!{\bf M}^{(n-1)}\left[(1-q)b_0,\frac{x-q}{1-q}\right]\bigg\} \;.
\end{split}
\ee
We can sum this into an integrodifferential equation where the ``time'' variable is $T$ instead of $\sigma$ used in~\eqref{integroDiffx},
\be\label{integroDiffConst}
\begin{split}
\frac{\partial{\bf M}}{\partial T}=&\int_0^1\ud q\bigg\{{\bf M}^L\!\cdot\!{\bf M}(b_0,x)\\
&+\theta(x-q){\bf M}^C\!\cdot\!{\bf M}\left[(1-q)b_0,\frac{x-q}{1-q}\right]\bigg\} \;,
\end{split}
\ee
with ``initial'' condition ${\bf M}(T=0)={\bf M}^{(0)}$. In contrast to~\eqref{integroDiffx}, where the physical result is only obtained by, in the end, setting $\sigma\to-\infty$, all values of $T$ in~\eqref{integroDiffConst} give physical results, and if we are interested in a field with e.g. $T=10$ then we obtain the results for $T<10$ as a byproduct because we integrate the equation starting with the initial condition at $T=0$.  

\begin{figure}
\includegraphics[width=\linewidth]{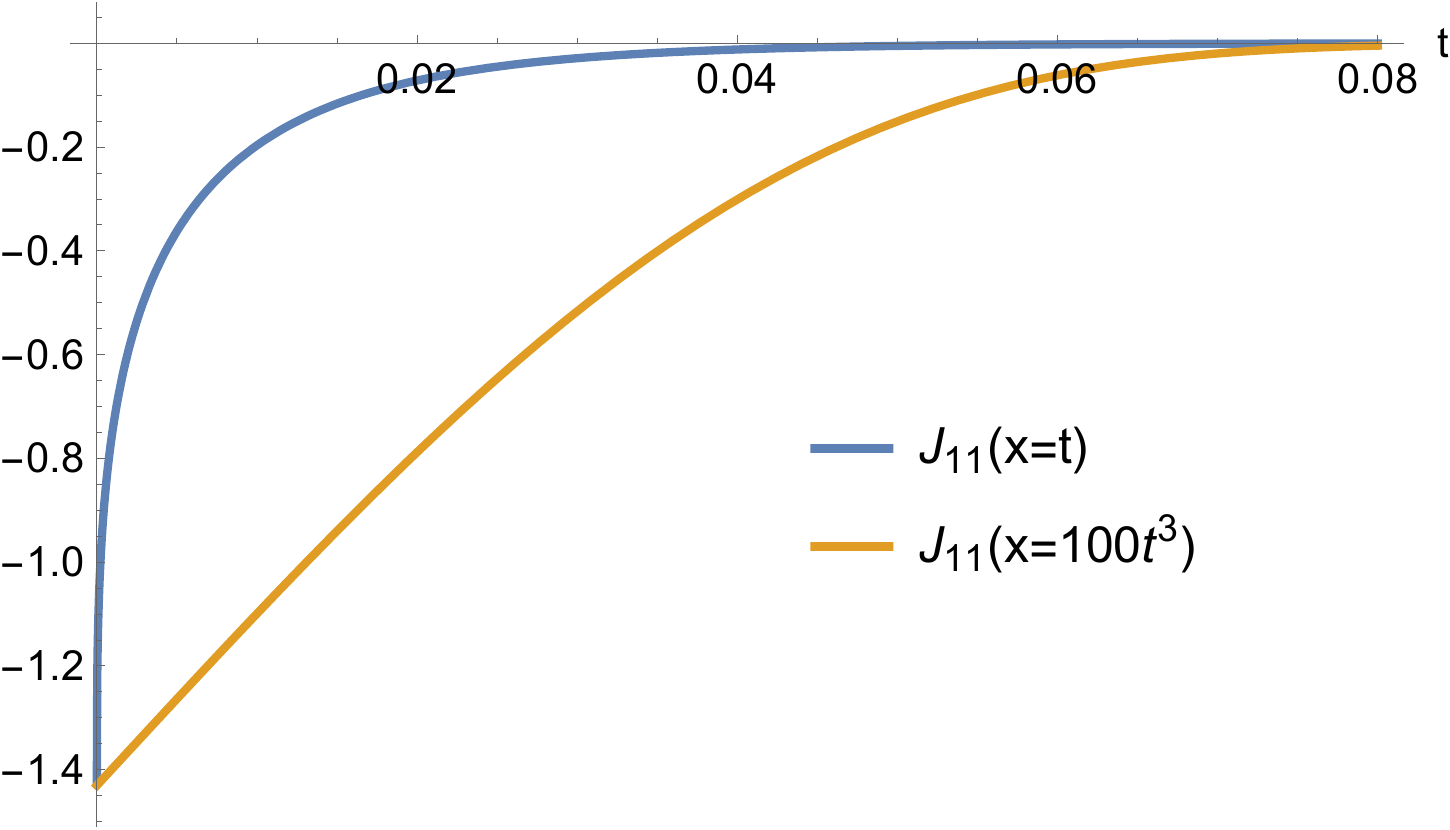}
\caption{Integral $\{1,0\}\cdot{\bf J}\cdot\{1,0\}$ in~\eqref{Jdef}, for $\chi=0.01$.}
\label{J11fig}
\end{figure}

The results for $\chi=0.01$ are shown in Fig.~\ref{Mchi0p01fig}. For these results we have used a step size of $\Delta T=0.1$. This is smaller than what it might first seem because the natural $\mathcal{O}(1)$ ``time'' variable is actually $\chi T$ (see~\eqref{ufromT}). Fig.~\ref{Mchi0p01fig} shows the results for every tenth time step (or every twentieth if one counts the midpoints). For $\chi$ we have used an evenly distributed grid from $\chi=0$ to $\chi=\chi_{\rm max}=0.01$ with $\Delta\chi=\chi_{\rm max}/100$. As can be seen in the plot for $M_{11}$ in Fig.~\ref{Mchi0p01fig}, $\partial{\bf M}/\partial x$ diverges at $x\to0$ for small $T$. Recall that this is because we have assumed a sharply peaked wave packet and is why we work with the cumulative distribution. While ${\bf M}$ is finite and all the integrals converge, we want to avoid having to use a large number of points in $x$ in order to obtain a good interpolation near $x=0$, because that would make the code much slower. For small $T$ we therefore use an evenly distributed grid in $y=x^{1/3}$ instead of $x$, which is better because $\partial{\bf M}/\partial y$ does not diverge at $y=0$ and is easier to interpolate using fewer points. 
We can illustrate this by looking at the part of the integral in~\eqref{integroDiffConst} with $q>x$,
\be\label{Jdef}
{\bf J}(b_0,x)=\int_x^1\!\ud q\,{\bf M}_L \;.
\ee
Since this is independent of ${\bf M}$, we have made a separate interpolation of it so that we do not have to calculate it over and over again when solving~\eqref{integroDiffConst}. Fig.~\ref{J11fig} shows that, as a function of $x$, $\partial{\bf J}/\partial x$ diverges as $x\to0$, but $\partial{\bf J}/\partial y$ is finite and therefore easier to interpolate. 
We have used $\Delta y=0.01$. As $T$ increases beyond a certain finite point, ${\bf M}$ becomes exponentially small near $x=0$, and then we no longer have the problem of a divergent $\partial{\bf M}/\partial x$. When this happens we switch to using an even grid in $x$ instead of $y$, with $\Delta x=0.01$. Thus at each point in $T$ we make a cubic-polynomial interpolation of the list of ${\bf M}(\chi,x)$ evaluated on a $100\times100$ size grid.  

\begin{figure}
\includegraphics[width=\linewidth]{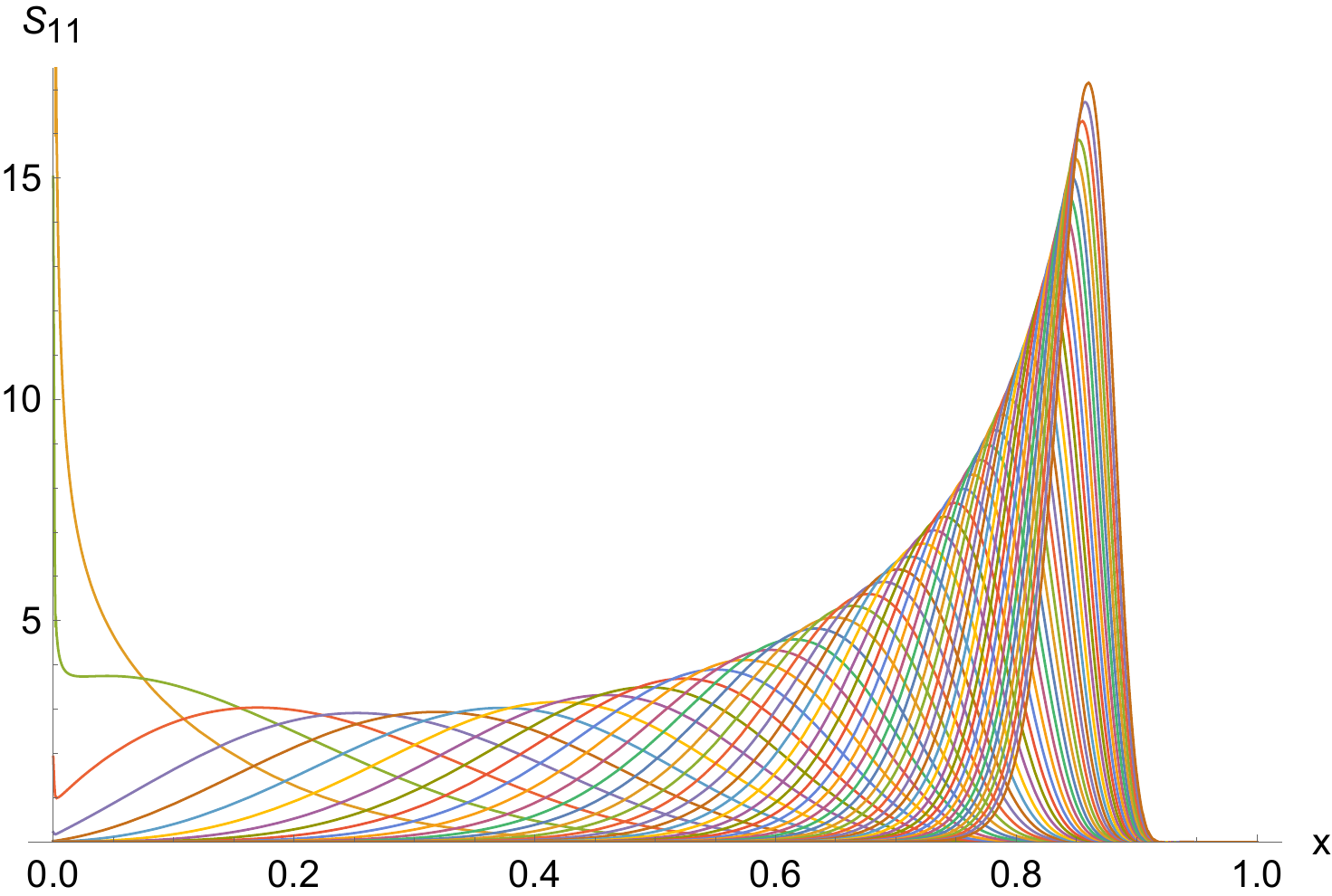}
\includegraphics[width=\linewidth]{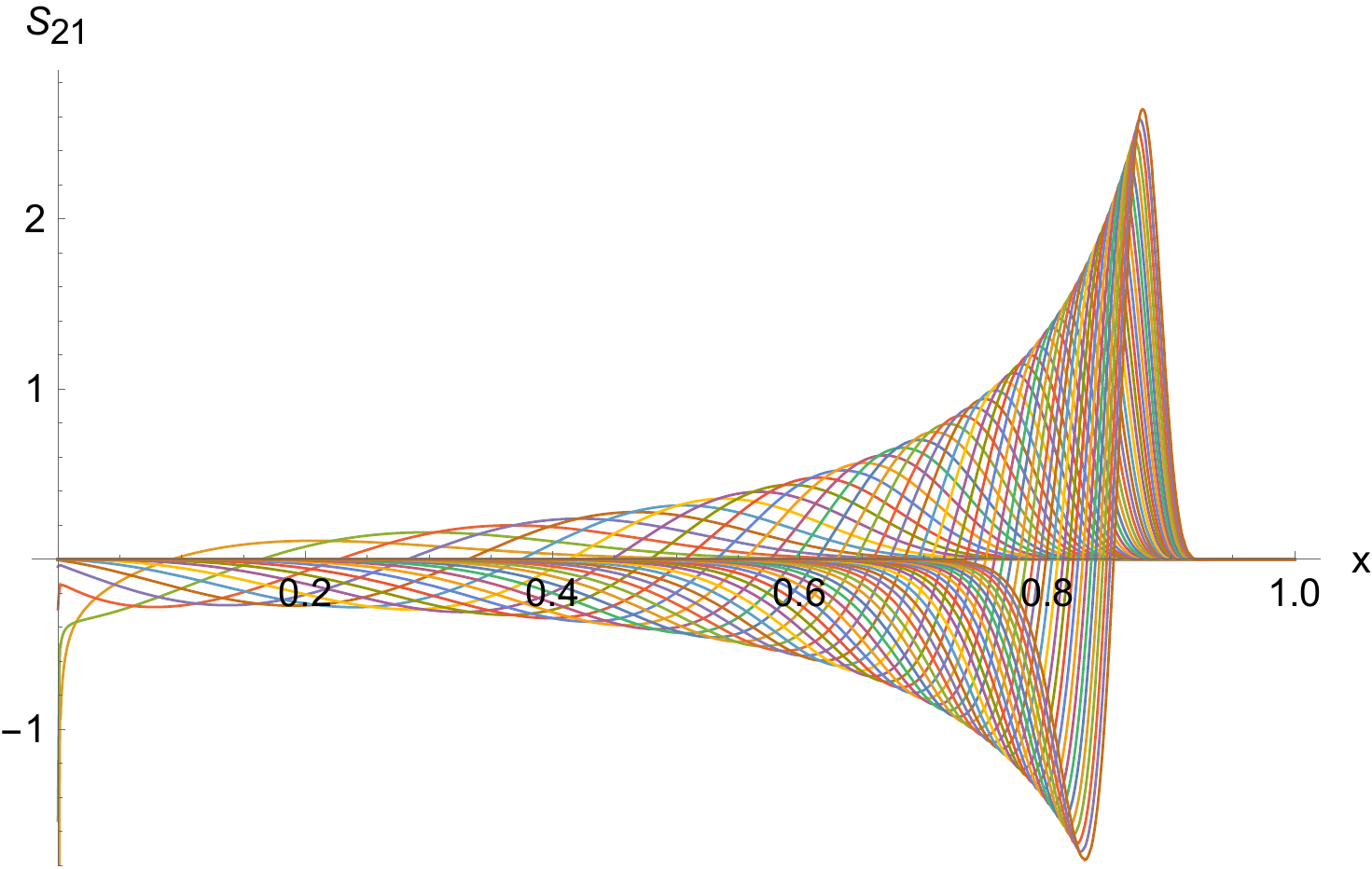}
\caption{Spectrum $S_{11}=\{1,0\}\cdot{\bf S}\cdot\{1,0\}$ and $S_{21}=\{0,1\}\cdot{\bf S}\cdot\{1,0\}$, for $\chi=0.1$ and for $T=0,2,4\dots1000$.}
\label{Schi0p1fig}
\end{figure}

The results for ${\bf S}$ for $\chi=0.1$ are shown in Fig.~\ref{Schi0p1fig}. The spectrum is initially a delta-function-like peak (because we hav neglected the width of the wave packet), then becomes a relatively low and wide Gaussian peak, but as $T$ increases further the peak
eventually starts to become higher and narrower as it is squeezed towards the upper limit $x=1$. This evolution of $\{1,0\}\cdot{\bf S}\cdot\{1,0\}$ is similar to the time evolution of distribution functions found in Fig.~1 in~\cite{Shen1972} and Fig.~2 in~\cite{Vranic:2015sft}, even though both methods and setups are different, e.g. we consider a single electron with a wave packet that is initially sharply peaked, while~\cite{Shen1972,Vranic:2015sft} considered electron beam distributions and neglected spin.

\subsection{Moments}

The moments\footnote{We have included a factor of $a_0$ here so that $a_0$ only appears in $\chi$ or $T$.} $\langle(a_0kP)^m\rangle$ can be obtained by solving  
\be\label{integroDiffConstMoments}
\begin{split}
\frac{\partial}{\partial T}\tilde{\bf M}(T,\chi)=&\int_0^1\ud q\bigg\{{\bf M}^L\!\cdot\!\tilde{\bf M}(T,\chi)\\
&+{\bf M}^C\!\cdot\!\tilde{\bf M}\left[T,(1-q)\chi\right]\bigg\} \;,
\end{split}
\ee
with ``initial'' condition 
\be
\tilde{\bf M}(T=0,\chi)=\chi^m{\bf 1} \;.
\ee
The derivation of this equation is a trivial generalization of the $m=0$ and $m=1$ cases in~\cite{Torgrimsson:2021wcj,Torgrimsson:2021zob}. Note that with this approach we obtain the moments without calculating the spectrum. We thus have two independent ways of calculating the moments: either by solving~\eqref{integroDiffConstMoments} or by first obtaining the spectrum via~\eqref{integroDiffConst} and then the moments by integrating as in~\eqref{MnFromIntM}.  

\begin{figure}
\includegraphics[width=\linewidth]{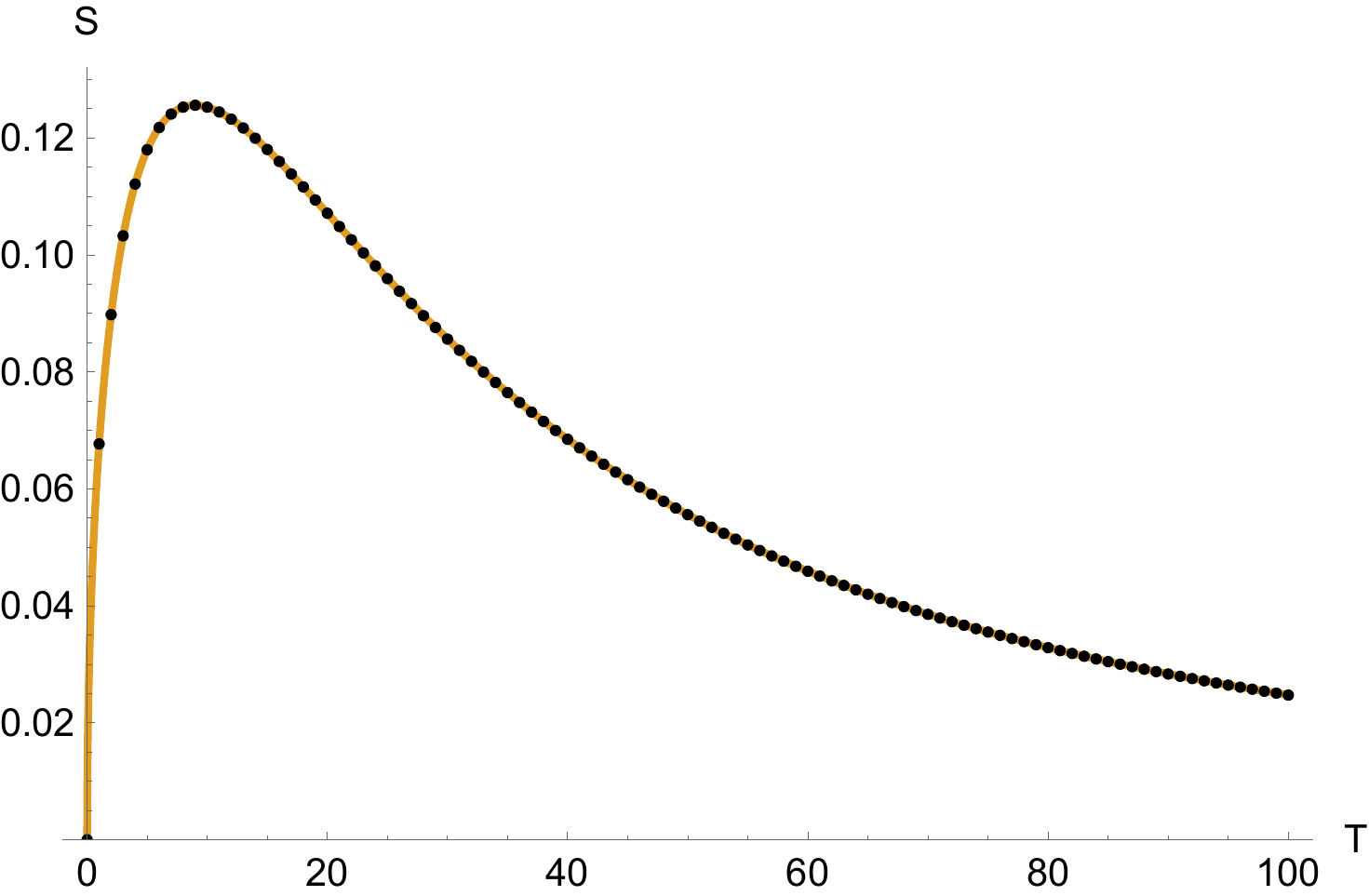}
\caption{The standard deviation~\eqref{standardDevDef} for $\chi=0.1$ and $\{1,0\}\cdot{\bf M}\cdot\{1,0\}$. The solid line is obtained using~\eqref{recursiveConstMoments} and the double resummation, and the dots are obtained by integrating ${\bf M}_{11}$ as in~\eqref{MnFromIntM} for every tenth time step. The corresponding $\partial_x{\bf M}$ is shown in Fig.~\ref{Schi0p1fig}.}
\label{standardFig}
\end{figure}

We have solved~\eqref{integroDiffConstMoments} numerically using the midpoint method for the variable $T$, as described in~\cite{Torgrimsson:2021zob}. At each step in $T$ we make an interpolation function for the $\chi$ dependence, with $0<\chi<\chi_{\rm max}$, where $\chi_{\rm max}$ is the maximum value of $\chi$ we want to consider. We have used an adaptive grid for $\chi$, i.e. adding points until a certain precision and accuracy is reached, which results in more points where the function has a larger curvature and less points where it is flat. The interpolation between these points are done using Mathematica's built-in function ``Interpolation'' with cubic polynomials.

$\tilde{\bf M}$ in \eqref{integroDiffConstMoments} already gives the resummation of all orders in $\alpha$. We can also obtain the moments by first calculating each order in $\alpha$ separately by solving the recursive formula
\be\label{recursiveConstMoments}
\begin{split}
\tilde{\bf M}^{(n)}(\chi)=\int_0^1\frac{\ud q}{n}&\bigg\{{\bf M}^L\!\cdot\!\tilde{\bf M}^{(n-1)}(\chi)\\
&+{\bf M}^C\!\cdot\!\tilde{\bf M}^{(n-1)}\left[(1-q)\chi\right]\bigg\} \;,
\end{split}
\ee
starting with
\be
\tilde{\bf M}^{(0)}(\chi)=\chi^m{\bf 1} \;,
\ee
and then resum the $\alpha$ expansion at the end.
Also in this case we can make a numerical interpolation function for the $\chi$ dependence for each step. After having obtained terms up to e.g. $n=10$ or $n=20$, we can resum the $\alpha$ expansion using Pad\'e approximants as described in~\cite{Torgrimsson:2021wcj,Torgrimsson:2021zob}. The convergence of this resummation method is often fast. Alternatively, we can expand each order in $\alpha$ as a power series in $\chi$. Then the $\chi$ expansion of $\tilde{\bf M}^{n}$ is obtained by inserting the $\chi$ expansion of $\tilde{\bf M}^{n-1}$ into~\eqref{recursiveConstMoments} and expanding the result in $\chi$. After having obtained $\chi$ expansions for the first e.g. $n=10$ orders in the $\alpha$ expansion, we first resum the $\chi$ expansions for each order in $\alpha$ separately using standard Borel-Pad\'e resummation, and finally we resum the $\alpha$ expansion using Pad\'e approximants (this last step is the same with or without using the $\chi$ expansion). A detailed explanation for this double-resummation approach can be found in~\cite{Torgrimsson:2021wcj,Torgrimsson:2021zob}. It turns out to be quite fast.

As an example, we consider $\chi=0.1$. The zeroth and first moments look very similar to the results already presented in~\cite{Torgrimsson:2021wcj,Torgrimsson:2021zob}. Calculating also the second moment allows us to obtain the standard deviation,
\be\label{standardDevDef}
S^2:=\frac{\langle(a_0 kP)^2\rangle}{\chi^2}-\left(\frac{\langle a_0 kP\rangle}{\chi}\right)^2 \;.
\ee 
In Fig.~\ref{standardFig} we plot $S$ obtained by using either the double resummation approach or by integrating the cumulative function as in~\eqref{MnFromIntM}. We find perfect agreement. Comparing with the double-resummation approach serves as a quick way to check the precision of ${\bf M}$(x), which takes much, much longer time to obtain.  
The shape of $S$ is quite similar to plots in Fig.~4 in~\cite{Vranic:2015sft}, Fig.~3 in~\cite{Ridgers2017} or Fig.~9 in~\cite{NielPRE2018} for the energy spread of an electron beam/bunch, i.e. first a rapid increase and then a slower decrease after a maximum. But, again, it should be noted that we are considering somewhat different quantities.

\subsection{Low energy expansion of moments}

We will now calculate the first quantum correction in a low energy expansion. $\chi\ll1$ is the expansion parameter, but we need to consider
\be\label{ufromT}
u:=\frac{2}{3}\chi T
\ee
as $\mathcal{O}(1)$ to keep a nontrivial dependence on $T$\footnote{There can however be experimental signals in the spectrum of the emitted photons even if $u\ll1$~\cite{King:2023avg}.}. The zeroth order is then given by
\be
\tilde{\bf M}=\frac{\chi^m}{(1+u)^m}{\bf 1} \;.
\ee
This is just the longitudinal-momentum component of the solution to LL~\cite{exactSolLL} to the power of $m$. The standard deviation~\eqref{standardDevDef}
vanishes at $\mathcal{O}(\chi^0)$, as it has to in the classical limit. To obtain the first nonzero correction we can solve the recursive equation~\eqref{recursiveConstMoments} approximately using the ansatz
\be\label{MomentsNLO}
\tilde{\bf M}=\chi^m\left(\frac{\bf 1}{(1+u)^m}+\delta\tilde{\bf M}\chi\right) \;,
\ee
where $\delta\tilde{\bf M}$ only depends on $m$ but not on $\chi$. With
\be
\delta\tilde{\bf M}=\sum_{n=0}^\infty u^n\delta\tilde{\bf M}^{(n)}
\ee
the problem has been reduced to an algebraic recursive equation for $\delta\tilde{\bf M}^{(n)}$ as function of $n$. There are standard methods to solve such recursive equations~\cite{BenderOrszag}, but this is most conveniently done using Mathematica's ``RSolve''. The result for $\delta\tilde{\bf M}^{(n)}$ is not particularly illuminating. But one thing to note though is that it is an expression valid for all orders in $\alpha$, which we can obtain here because we only consider the first two orders in the $\chi$ expansion. Contrast this with the general case described above where we would only be able to calculate the first e.g. $10$ or $20$ orders and then use e.g. Pad\'e resummation based on those finite number of terms. Now for the low-energy approximation, we have access to all coefficients in the $\alpha$ expansion and we can resum this using Mathematica, for example. For the average momentum, i.e. $m=1$, we find
\be\label{MnloAve}
\begin{split}
&\delta\tilde{\bf M}=\\
&\begin{pmatrix}\frac{55[u+2(1+u)\ln(1+u)]}{16\sqrt{3}(1+u)^3} & \frac{3[u-\ln(1+u)]}{2(1+u)^2} \\
-\frac{3\ln(1+u)}{2(1+u)^2} & \frac{5[2u-9u^2+22(1+u)\ln(1+u)]}{16\sqrt{3}(1+u)^3}\end{pmatrix} \;.
\end{split}
\ee
We recognize the $\{0,1\}\cdot\delta\tilde{\bf M}\cdot\{1,0\}$ element from Eq.~(13) in~\cite{Torgrimsson:2021wcj}, which corresponds to the difference in the final momentum due to initial spin being either parallel or antiparallel to the magnetic field and after summing over the final spins. For $m=2$ we find
\be\label{MnloSq}
\begin{split}
&\delta\tilde{\bf M}=\\
&\begin{pmatrix}\frac{55[3u+4(1+u)\ln(1+u)]}{16\sqrt{3}(1+u)^4} &\frac{3[u-2\ln(1+u)]}{2(1+u)^3} \\-\frac{3\ln(1+u)}{(1+u)^3} &\frac{5[24u-9u^2+44(1+u)\ln(1+u)]}{16\sqrt{3}(1+u)^4} \end{pmatrix} \;.
\end{split}
\ee
If we sum over the final spin then we find a standard deviation that to leading order does not depend on the initial spin,
\be\label{standardLead}
S^2=\frac{55u}{16\sqrt{3}(1+u)^4}\chi+\mathcal{O}(\chi^2) \;.
\ee
Thus, the width of the spectrum goes to zero at both $u\ll1$ and $u\gg1$, which is also what we see in Fig.~\ref{Schi0p1fig}.
If we expect the spectrum to have a more or less symmetric peak around the average momentum, then it is not surprising that $S\to0$ asymptotically, because the average momentum decreases as $1/(1+u)$ and the lightfront longitudinal momentum $kP>0$, so the lower half of the peak will be squeezed between $1/(1+u)$ and $0$, and hence the peak must become narrower. However, even if we normalize the width by dividing $S$ by $1/(1+u)$ the result still goes to zero. In~\cite{Torgrimsson:2021wcj} we showed that the average momentum converges to the classical LL momentum asymptotically, and now we can also see that the standard deviation decreases asymptotically. 

We also note that the width scales as $S\sim\sqrt{\chi}$. So, if, say, $\chi=0.01$ then we would expect $S\sim0.1$, which corresponds to $10\%$ of the physical interval for this scaled momentum variable (i.e. $0$ to $1$). Thus, even if $\chi$ is so small that there is little difference between the average momentum and its classical limit (the solution to LL), the width can still be a significant fraction. 

While it would be possible to extend the above approach to include both higher powers of $\chi$ as well as higher moments, it becomes inconvenient to solve the recursive equations for the expansions in $u$ and then resumming them, so we have instead used the following approach for higher orders. We change variable in~\eqref{integroDiffConstMoments} from $T$ to $u$ in~\eqref{ufromT}. We separate out the overall factor of $\chi^m$ in~\eqref{MomentsNLO} as
\be\label{tildeMtoW}
\tilde{\bf M}(T,\chi,m)=\chi^m{\bf W}(u,\chi,m) \;.
\ee
The integrodifferential equation in terms of ${\bf W}$ is given by
\be\label{integroDiffConstMomentsW}
\begin{split}
\frac{\partial}{\partial u}{\bf W}(u,\chi)&=\frac{3}{2}\int_0^1\frac{\ud q}{\chi}\bigg\{{\bf M}^L\!\cdot\!{\bf W}(u,\chi)\\
&+(1-q)^m{\bf M}^C\!\cdot\!{\bf W}\left[(1-q)u,(1-q)\chi\right]\bigg\} \;,
\end{split}
\ee 
with the same ``initial'' condition
\be
{\bf W}(u=0,\chi,m)={\bf 1}
\ee
for all moments. Note that, while~\eqref{integroDiffConstMoments} is local in $T$, \eqref{integroDiffConstMomentsW} is not local in $u$. The reason for making this change of variables is to have a natural $\chi$ expansion right from the start and to avoid expanding in $u$. This expansion can now be written as
\be\label{Wtow}
{\bf W}(u,\chi,m)=\sum_{k=0}^\infty{\bf w}_{m,k}(u)\chi^k \;.
\ee
To find the first couple of ${\bf w}$'s, we insert~\eqref{Wtow} into~\eqref{integroDiffConstMomentsW} and match the two sides order by order in $\chi$. To do so we need a suitable change of variable for the $q$ integral. We have used $\gamma$ as defined by
\be\label{qFromgamma}
q=\frac{\chi\gamma}{1+\chi\gamma} \;.
\ee
After this change of variable, we can expand the integrand before performing the integral. This results in the following type of integrals,
\be
\begin{split}
&\int_0^\infty\ud\gamma\,\gamma^n\frac{\text{Ai}(\gamma^{2/3})}{\gamma^{1/3}}
=\frac{3^{\frac{1}{2}+n}}{4\pi}\Gamma\left[\frac{1}{3}+\frac{n}{2}\right]\Gamma\left[\frac{2}{3}+\frac{n}{2}\right] \\
&=\left\{\frac{1}{2},\frac{\sqrt{3}}{4},1,\frac{35 \sqrt{3}}{16},20,\frac{5005
   \sqrt{3}}{64},\dots\right\} \;,
\end{split}
\ee
where the second line shows the explicit numbers for $n=0,1,2\dots$,
\be
\begin{split}
&\int_0^\infty\ud\gamma\,\gamma^n\frac{\text{Ai}'(\gamma^{2/3})}{\gamma^{2/3}}
=-\frac{3^{\frac{1}{2}+n}}{4\pi}\Gamma\left[\frac{1}{6}+\frac{n}{2}\right]\Gamma\left[\frac{5}{6}+\frac{n}{2}\right]\\
&=\left\{-\frac{\sqrt{3}}{2},-\frac{1}{2},-\frac{5 \sqrt{3}}{8},-4,-\frac{385
   \sqrt{3}}{32},-140,\dots\right\} \;,
\end{split}
\ee
\be
\begin{split}
&\int_0^\infty\ud\gamma\,\gamma^n\text{Ai}_1(\gamma^{2/3})
=\frac{3^{\frac{1}{2}+n}}{2\pi(1+n)}\Gamma\left[\frac{5}{6}+\frac{n}{2}\right]\Gamma\left[\frac{7}{6}+\frac{n}{2}\right]\\
&=\left\{\frac{1}{2
   \sqrt{3}},\frac{1}{3},\frac{35}{24
   \sqrt{3}},\frac{10}{3},\frac{1001}{32
   \sqrt{3}},\frac{1120}{9},\dots\right\} \;.
\end{split}
\ee
These are the same integrals that we used in~\cite{Torgrimsson:2021wcj} to calculate the $\chi$ expansion of each of the orders in $\alpha$ separately. Now we instead work with~\eqref{Wtow}, which is already resummed in $\alpha$ (recall $u\propto\alpha$).

To zeroth order we find
\be
{\bf w}_{m,0}'(u)=-\frac{m}{1+u}{\bf w}_{m,0}(u) \;,
\ee
which implies
\be
{\bf w}_{m,0}(u)=\frac{\bf 1}{(1+u)^m} \;,
\ee
in agreement with~\eqref{MomentsNLO} and what one should expect from the solution to LL.
${\bf w}_{m,k}(u)$ starts contributing at $\mathcal{O}(\chi^k)$ in the expansion of~\eqref{integroDiffConstMomentsW}. At this order, ${\bf w}_{m,0}(u)$ to ${\bf w}_{m,k-1}(u)$ also contribute. Moving all the ${\bf w}_{m,k}(u)$ terms to the left-hand side gives a differential equation on the following form
\be\label{dwmk}
\begin{split}
\frac{1}{(1+u)^{m+k-1}}&\frac{\ud}{\ud u}[(1+u)^{m+k}{\bf w}_{m,k}(u)]\\
&=F[{\bf w}_{m,0},...,{\bf w}_{m,k-1}] \;.
\end{split}
\ee  
We find that ${\bf w}_{m,k}(u)$ can be expressed in terms of linear combinations of
\be
\mathcal{L}_{r,s}(u):=\frac{\ln^r(1+u)}{(1+u)^s} \;,
\ee
where $r$ and $s$ are integers. Derivatives of $\mathcal{L}_{r,s}$ appear when we replace $u\to(1-q)u$, change variable to $\gamma$ and expand in $\chi$, but
\be
\mathcal{L}_{r,s}'(u)=r\mathcal{L}_{r-1,s+1}(u)-s \mathcal{L}_{r,s+1}(u)
\ee
so this class of functions is closed under differentiation. We can integrate~\eqref{dwmk} using
\be\label{LrsInt}
\begin{split}
\int_0^u\ud u'\,&\mathcal{L}_{r,s}(u')=\frac{r!}{(s-1)^{r+1}}\\
&-\frac{1}{(1+u)^{s-1}}\sum_{j=0}^r\frac{r!}{(r-j)!}\frac{\ln^{r-j}(1+u)}{(s-1)^{j+1}} 
\end{split}
\ee 
and\footnote{\eqref{Lr1int} can be obtained from~\eqref{LrsInt} by treating $s=1+\delta$ as a continuous variable and expanding in $|\delta|\ll1$.}
\be\label{Lr1int}
\int_0^u\ud u'\,\mathcal{L}_{r,1}(u')=\frac{\ln^{1+r}(1+u)}{1+r} \;,
\ee
so this class of functions is also closed under integration. 

For $m=0$ we find
\be
{\bf w}_{0,1}=\frac{3}{2}\frac{u}{1+u}\begin{pmatrix}0&1\\0&-\frac{5\sqrt{3}}{8}\end{pmatrix} \;,
\ee
which agrees with Eq.~(10) in~\cite{Torgrimsson:2021zob}, and for $m=1$ and $m=2$ we recover~\eqref{MnloAve} and~\eqref{MnloSq}. At the next order in $\chi$ we find, for example,
\be
\begin{split}
\{1,0\}\cdot&{\bf w}_{1,2}\!\cdot\!\{1,0\}=\\
&-\frac{u \left(22656+15736 u+2155
   u^2\right)}{768
   (1+u)^5}\\
&+\frac{\left(-2368+369 u-288
   u^2\right) \ln(1+u)}{128
   (1+u)^4}\\
&+\frac{3457 \ln^2(1+u)}{192
   (1+u)^3} \;.
\end{split}
\ee
As we go to even higher orders we get more and more terms, and the expressions become too large to write down here. We will use them, though, in the next section to determine integration constants in the derivation of an analytical approximation of the spectrum.

\subsection{Low energy approximation of the spectrum}

\begin{figure}
\includegraphics[width=\linewidth]{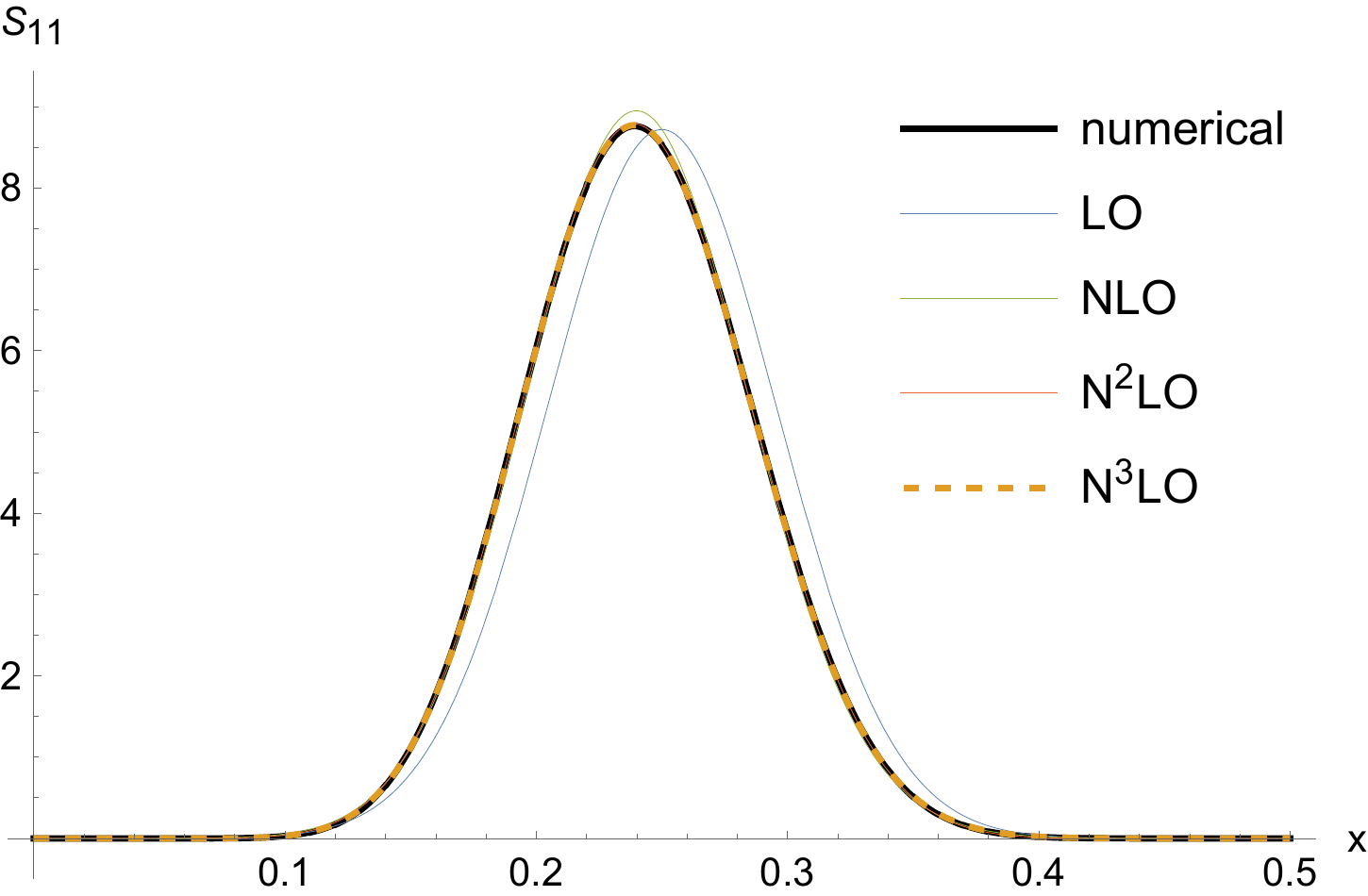}
\includegraphics[width=\linewidth]{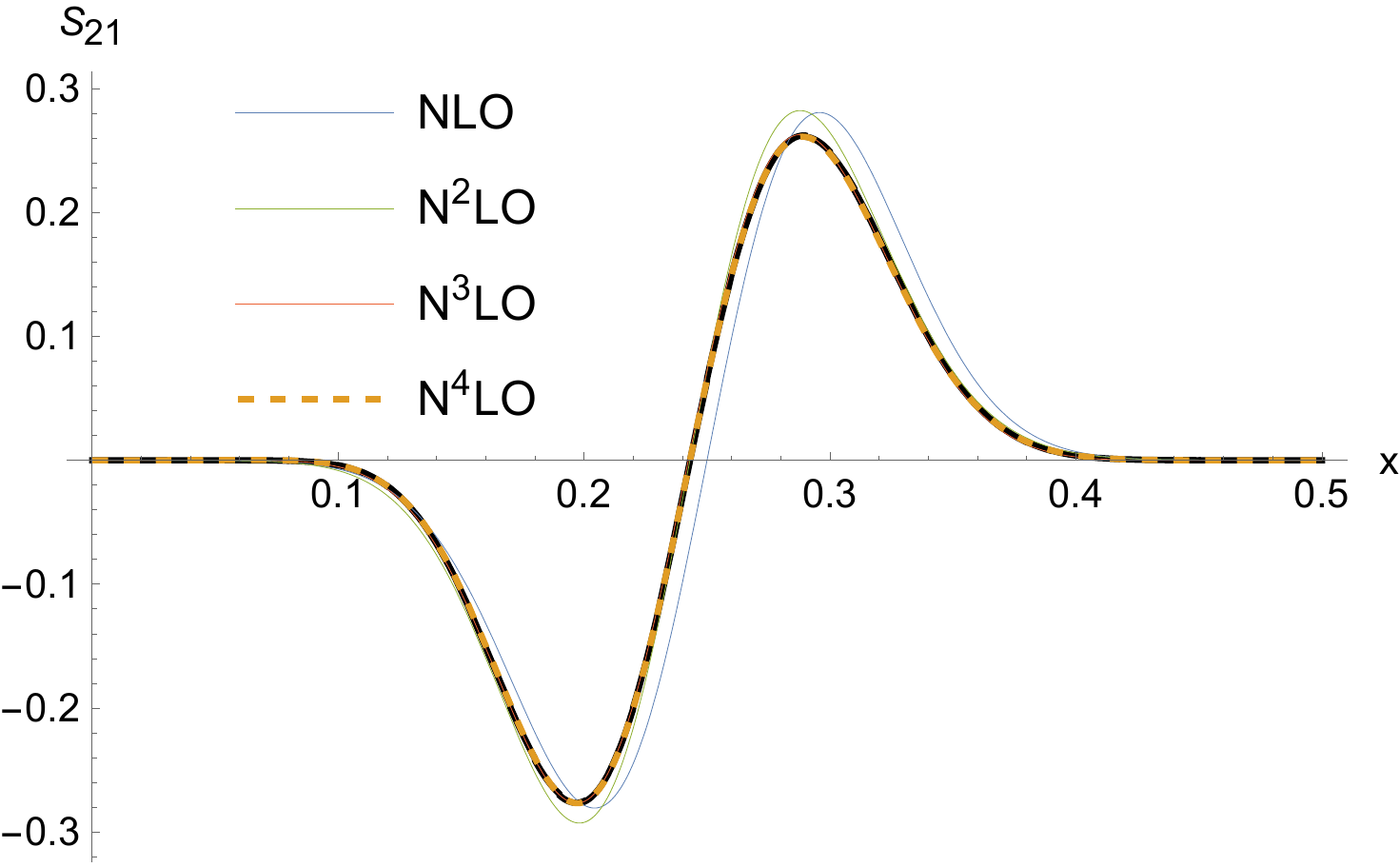}
\caption{Comparison of the approximation in~\eqref{Sansatz} and a numerical solution to~\eqref{integroDiffConst}, for $\chi=0.01$ and $T=50$. LO refers to the $n=0$ term~\eqref{Sansatz} only, and N$^k$LO is the sum up to $n=k$. $S_{11}=\{1,0\}\cdot{\bf S}\cdot\{1,0\}$ and $S_{21}=\{0,1\}\cdot{\bf S}\cdot\{1,0\}$. $S_{21}$ is identically zero at LO. N$^2$LO and N$^3$LO for $S_{11}$, and N$^3$LO and N$^4$LO for $S_{21}$, are basically indistinguishable on the scale of these plots.}
\label{SapproxFig}
\end{figure}

To obtain an analytical approximation for the spectrum, we start by differentiating~\eqref{integroDiffConst} with respect to $x$ to obtain an equation directly expressed in terms of ${\bf S}$, 
\be\label{integroDiffConstSpec}
\begin{split}
\frac{\partial{\bf S}}{\partial T}=&\int_0^1\ud q\bigg\{{\bf M}^L\!\cdot\!{\bf S}(\chi,x)\\
&+\frac{\theta(x-q)}{1-q}{\bf M}^C\!\cdot\!{\bf S}\left[(1-q)\chi,\frac{x-q}{1-q}\right]\bigg\}\\
&+{\bf M}_C(\chi,q=x)\!\cdot\!{\bf M}[(1-x)\chi,0] \;.
\end{split}
\ee
The last term makes this an inhomogeneous equation. It is given by the time-ordered exponential in~\eqref{timeOrderedExpx0}. However, for low energies it is exponentially suppressed and can be neglected. 

We know from~\eqref{standardLead} that the width of the peak in the spectrum is $\mathcal{O}(\sqrt{\chi})$. In the limit $\chi\to0$ this would be an increasingly narrow peak, centered at a point which we denote $x=x_{cl}(u)$, where we again use $u$ in~\eqref{ufromT}. In the classical limit, we have $\langle kP\rangle/kp\to1/(1+u)$ (the solution to LL), so from~\eqref{xDefinition} we have
\be\label{xcl}
x_{cl}(u)=\frac{u}{1+u} \;.
\ee
If we were to plot ${\bf S}(x)$ for very small $\chi$, it would be natural to not plot it over the entire interval $0<x<1$, because then we would just see a very sharp peak. Instead we would plot it over $x_{cl}-c_1\sqrt{\chi}<x<x_{cl}+c_2\sqrt{\chi}$, where $c_1$ and $c_2$ are roughly $\sim 5-10$ or so. Beyond this interval ${\bf S}(x)$ would anyway be negligible small. Similarly, in order to obtain an analytic approximation, we also want to figure out how to choose parameters such that they can be considered $\mathcal{O}(\chi^0)$. Here we keep $u$ as ``time'' variable, but switch to
\be\label{Xfromx}
X=\frac{x-x_{cl}(u)}{\sqrt{\chi\lambda(u)}}
\ee
as a momentum parameter. $\lambda(u)$ is related to the standard deviation and will be determined below. As an educated guess, and after some trial and error, we take as an ansatz
\be\label{Sansatz}
{\bf S}=\frac{1}{\sqrt{\chi\lambda}}\mathcal{E}(X^2)\sum_{n=0}^\infty{\bm\rho}_n(u,X)\chi^{n/2} \;.
\ee
The idea now is to insert this expansion into~\eqref{integroDiffConstSpec} to determine the functions $\lambda$, $\mathcal{E}$ and ${\bm\rho}$. We again change integration variable from $q$ to $\gamma$ as in~\eqref{qFromgamma}, which allows us to expand the integrand in a series in $\chi$ before performing the integral. This expansion is done with $u$ and $X$ considered as $\mathcal{O}(1)$ parameters. In particular, for the ${\bf M}_C\cdot{\bf S}$ term, the replacements $\chi\to(1-q)\chi$ and $x\to(x-q)/(1-q)$ lead to
\be
\begin{split}
X\to& X+(x_{cl}+ux'_{cl}-1)\frac{\gamma\chi^{1/2}}{\sqrt{\lambda}}\\
&+\left(3+u\frac{\lambda'}{\lambda}\right)\frac{X\gamma\chi}{2}+\mathcal{O}(\chi^{3/2}) \;.
\end{split}
\ee 
With
\be
\frac{\partial}{\partial T}{\bf S}(T,x)=\frac{2\chi}{3}\left(\frac{\partial}{\partial u}+\frac{\partial X}{\partial u}\frac{\partial}{\partial X}\right){\bf S}(u,X) 
\ee
and
\be
\frac{\partial X}{\partial u}=-\frac{x_{cl}'}{\sqrt{\chi\lambda}}-\frac{X\lambda'}{2\lambda}
\ee
we see that the expansion of $\partial{\bf S}/\partial T$ and hence also the right-hand side of~\eqref{integroDiffConstSpec} starts at $\mathcal{O}(\chi^0)$. 

Rearranging~\eqref{integroDiffConstSpec} as RHS$-$LHS$=0$, we find at $\mathcal{O}(\chi^0)$
\be\label{RHSmLHSlead}
\frac{2}{3\lambda}[(1+u)x_{cl}'+x_{cl}-1][2X\mathcal{E}'{\bm\rho}_0+\mathcal{E}\frac{\partial{\bm\rho}_0}{\partial X}]=0 \;.
\ee
With initial condition $x_{cl}(0)=0$, \eqref{RHSmLHSlead} implies that $x_{cl}$ is given by~\eqref{xcl}.

At $\mathcal{O}(\chi^{1/2})$ we would in general get derivatives on ${\bm\rho}_0$, but as part of the ansatz we take ${\bm\rho}_0={\bf 1}$. We then find
\be\label{RHSmLHSnlo}
\begin{split}
&\bigg\{\frac{1}{\sqrt{\lambda}}\left[1+(1+u)\frac{\lambda'}{3\lambda}\right][\mathcal{E}+2X^2\mathcal{E}']\\
&+\frac{55}{24\sqrt{3}\lambda^{3/2}(1+u)^4}[\mathcal{E}'+2X^2\mathcal{E}'']\bigg\}{\bf 1}=0
\end{split} \;.
\ee
This equation should hold for any values of $u$ and $X$, and $\lambda$ ($\mathcal{E}$) should only depend on $u$ ($X$). If we first set $X=0$, then we obtain a differential equation for $\lambda$, which we solve with initial condition $\lambda(0)=0$, which is motivated by the fact that the standard deviation should go to zero at $u=0$, as we found in~\eqref{standardLead}. We find
\be\label{lambdaSol0}
\lambda(u)=-\frac{\mathcal{E}'(0)}{\mathcal{E}(0)}\frac{55u}{8\sqrt{3}(1+u)^4} \;.
\ee
Inserting~\eqref{lambdaSol0} into~\eqref{RHSmLHSnlo} gives and equation that only involves $X$,
\be
X^2\mathcal{E}''(X^2)+\left(\frac{1}{2}-\frac{\mathcal{E}'(0)}{\mathcal{E}(0)}X^2\right)\mathcal{E}'(X^2)-\frac{\mathcal{E}'(0)}{2\mathcal{E}(0)}\mathcal{E}(X^2)=0 \;.
\ee 
The solution to this equation is
\be
\mathcal{E}(X^2)=\mathcal{E}(0)\exp\left(\frac{\mathcal{E}'(0)}{\mathcal{E}(0)}X^2\right) \;.
\ee
At first it might seem like we have two integration constants to determine, $\mathcal{E}(0)$ and $\mathcal{E}'(0)$. However, from~\eqref{Xfromx} and~\eqref{lambdaSol0} we see that the constants in the exponent actually cancel. This is not surprising since we could have included any $\mathcal{O}(1)$ constant in the definition of $X$. We can therefore set $\mathcal{E}'(0)=-\mathcal{E}(0)$ so that the exponent is simply $e^{-X^2}$. The remaining constant is determined by
\be
\int_0^1\ud x\,{\bf S}={\bf M}(x=1)={\bf 1} \;,
\ee   
which to leading order implies
\be
\int_{-\infty}^\infty\ud X\,\mathcal{E}(X^2)=1 \;.
\ee
Thus, expanding~\eqref{integroDiffConstSpec} to next-to-leading order (i.e. $\mathcal{O}(\chi^{1/2})$ has allowed us to determined ${\bf S}$ to leading order, and we find that the width of the peak is determined by
\be\label{lambdaSol}
\lambda(u)=\frac{55u}{8\sqrt{3}(1+u)^4} 
\ee
and the shape of the peak by
\be\label{Esol}
\mathcal{E}(X^2)=\frac{1}{\sqrt{\pi}}e^{-X^2} \;.
\ee

However, since the expansion parameter in~\eqref{Sansatz} is only $\sqrt{\chi}$ rather than e.g. $\chi$, we expect the next couple orders to be important to obtain a precise approximation even for $\chi\sim0.01$, so we continue.

At $\mathcal{O}(\chi)$ we find a differential equation for ${\bm\rho}_1(u,X)$. This equation contains terms that are linear in ${\bm\rho}_1(u,X)$ and derivatives of ${\bm\rho}_1(u,X)$ with respect to $u$ and $X$, and terms without ${\bm\rho}_1(u,X)$. After dividing away the overall factor of $e^{-X^2}$, the terms without ${\bm\rho}_1(u,X)$ can be written $f_1(u)X+f_3(u)X^3$. The solution is therefore on the form
\be
{\bm\rho}_1(u,X)={\bm\rho}_{1,1}(u)X+{\bm\rho}_{1,3}(u)X^3 \;.
\ee
The equation for ${\bm\rho}_1$ thus separates into one part proportional to $X$ and another part proportional to $X^3$. Both parts have to be separately zero. Solving these equations gives
\begin{widetext}
\be\label{rho11}
{\bm\rho}_{1,1}(u)=\frac{1}{\sqrt{u}}\begin{pmatrix}\frac{c_1}{u}+c_2-\frac{\sqrt{55}}{\sqrt{2}3^{1/4}}\left[\frac{1}{1+u}+\ln(1+u)\right] & c_3+\frac{6\sqrt{2}3^{1/4}}{\sqrt{55}}\ln(1+u)\\
c_4+\frac{6\sqrt{2}3^{1/4}}{\sqrt{55}}\ln(1+u)&\frac{c_5}{u}+c_6-\frac{\sqrt{55}}{\sqrt{2}3^{1/4}}\left[\frac{1}{1+u}+\ln(1+u)\right] \end{pmatrix}
\ee
and
\be\label{rho13}
{\bm\rho}_{1,3}(u)=\frac{1}{\sqrt{u}}\begin{pmatrix}-\frac{2c_1(1+u)}{3u}+\frac{\sqrt{55}}{\sqrt{2}3^{1/4}}\left[\frac{1}{1+u}-\frac{2129}{3025}\right]&0\\0& -\frac{2c_5(1+u)}{3u}+\frac{\sqrt{55}}{\sqrt{2}3^{1/4}}\left[\frac{1}{1+u}-\frac{2129}{3025}\right] 
\end{pmatrix} \;,
\ee
\end{widetext}
where $c_i$ are some constants. We cannot determine these constants by considering some initial conditions at $u=0$, because our approximation breaks down when $u$ becomes too small. We can see this from Fig.~\ref{Mchi0p01fig}, where the $x$ derivative diverges at $x=0$ for $T$ below some finite point. A diverging ${\bf S}$ is of course not approximated by~\eqref{Sansatz}. 
To determine the constants $c_i$ we will instead compare with the moments calculated with the method in the previous section.

We will use the same method also for higher orders, and we have found that ${\bm\rho}_n(u,X)$ is in general a polynomial in $X$. To compare with the moments we consider
\be\label{matchMoments}
\begin{split}
&\int_{-\infty}^\infty\frac{\ud X}{\sqrt{\pi}}X^m e^{-X^2}\sum_{n=0}^\infty{\bm\rho}_n(u,X)\chi^{n/2}=\int\ud x\,{\bf S} X^m\\
&=\frac{1}{(\chi\lambda)^{m/2}}\int\ud x\,{\bf S}\left(\frac{1}{1+u}-(1-x)\right)^m\\
&=\frac{1}{(\chi\lambda)^{m/2}}\sum_{k=0}^m\binom{m}{k}\frac{(-1)^k}{(1+u)^{m-k}}\sum_{l=0}^\infty{\bf w}_{k,l}\chi^l \;,
\end{split}
\ee
with ${\bf w}$ defined in~\eqref{Wtow}. By matching the first and the last lines for each $m$ and for each order in $\chi$ we obtain equations that we can use to determine the constants in~\eqref{rho11} and~\eqref{rho13}, and other constants at higher orders. 

By evaluating~\eqref{matchMoments} with $m=1$ and $m=3$ and selecting the term proportional to $\chi^{1/2}$ we are able to determine the constants in~\eqref{rho11} and~\eqref{rho13} as $c_1=c_3=c_4=c_5=0$ and
\be
c_2=c_6=\frac{1681}{55\sqrt{110}3^{1/4}} \;.
\ee 

In Fig.~\ref{SapproxFig} we compare the approximation in~\eqref{Sansatz} with a numerical solution of~\eqref{integroDiffConst}. We have chosen $\chi=0.01$, which at first might seem like a quite small value. However, the approximation is a series in $\sqrt{\chi}=0.1$, so we should not expect higher orders to be negligible. In Fig.~\ref{SapproxFig} we see that, while the leading order gives a decent approximation, adding higher orders does indeed lead to a noticeable difference. Even after adding the next-to-leading order there is still a noticeable difference from the numerical result. For $\{1,0\}\cdot{\bf S}\cdot\{1,0\}$, it is only after adding the first three terms that we obtain a result that is more or less indistinguishable from the numerical solution. For $\{0,1\}\cdot{\bf S}\cdot\{1,0\}$ we need the first four terms.

We have mostly considered the cumulative function as a computationally convenient tool for in the end finding the spectrum, but to obtain this low-energy approximation of the spectrum we worked directly with ${\bf S}$. If we want ${\bf M}$ it is now straightforward to integrate~\eqref{Sansatz}. We have found that the ${\bm\rho}$ functions are in general polynomials in $X$. Regardless of what sort of polynomial they are, we can always write them as sums of Hermite polynomials, $H_m(X)$. We can therefore write~\eqref{Sansatz} as
\be
{\bf S}=\frac{e^{-X^2}}{\sqrt{\chi\lambda\pi}}\sum_{m=0}^\infty{\bf h}_m(u,\chi)H_m(X) \;.
\ee 
This is useful because integrating over $X$ becomes trivial using Rodrigues' formula,
\be
H_m(X)=(-1)^m e^{X^2}\partial_X^m e^{-X^2} \;,
\ee
and we find
\be
\begin{split}
{\bf M}&=\int_{-\infty}^x\ud x'{\bf S}(x')\\
&=\frac{1}{2}[1+\text{erf}(X)]{\bf1}-\frac{e^{-X^2}}{\sqrt{\pi}}\sum_{m=1}^\infty{\bf h}_mH_{m-1}(X) \;.
\end{split}
\ee

\section{Finite wave packets}\label{Finite wave packets}

So far we have focused on sharply peaked wave packets. Now we consider wave packets with finite width. In principle we could consider a wave packet with two different functions, $f_1(p)$ and $f_2(p)$, for two different spin states, but we will for simplicity consider just a single function $f$. When we considered a wave packet sharply peaked at $b_0$, we found it natural to factor out $b_0$ as in~\eqref{xDefinition}. But now when we have an integral over the initial momentum, we replace $(1-x)b_0\to b'$, so that we can describe the final momentum without referring to the initial momentum. To avoid confusion of the integration variable $kp'$ with $b'$ we also replace as $kp'\to kp_{\rm out}$.
To generalize we first note that each contribution to the cumulative function involves momentum integrals on the form
\be
\begin{split}
&\int\ud\tilde{p}_{\rm out}\theta(kp_{\rm out}-b')\bigg|\int\tilde{p}_{\rm in}f(p_{\rm in})\\
&\times\frac{1}{k_\LCp}(2\pi)^3\delta_{\LCm,\LCperp}^3(p_{\rm out}+\sum_j l_{(j)}-p_{\rm in})M\bigg|^2\\
=&\int\ud\tilde{p}_{\rm in}|f(p_{\rm in})|^2\theta(kp_{\rm out}-b')\frac{|M|^2}{kp_{\rm in}kp_{\rm out}} \;,
\end{split}
\ee 
where in the last line
\be
p^{\rm out}_{\LCm,\LCperp}=p^{\rm in}_{\LCm,\LCperp}-\sum_j l^{(j)}_{\LCm,\LCperp} \;,
\ee
the sum is over all real photons emitted in this particular term, and the factor of $1/k_\LCp$ is just due to our normalization of the amplitude $M$. In this paper we only consider observables that do not depend on the transverse momenta. After integrating over the transverse momenta of the final-state particles, the probabilities in plane waves no longer depend on the transverse momentum of the initial particle. We write
\be
\int\ud\tilde{p}|f(p)|^2 h(kp)=\int_0^\infty\ud(kp)\rho(kp) h(kp) \;,
\ee
where $\rho$ now describes the initial longitudinal momentum distribution.
We thus find a simple relation between the Mueller matrix ${\bf M}(f;b')$ for a wide wave packet and ${\bf M}(b_0,b')$ for a sharply peaked wave packet\footnote{A similar incoherent relation for nonlinear Compton scattering in plane waves at $\mathcal{O}(\alpha)$ has been used in~\cite{wavePacketPeatross,Angioi:2016vir} to study effects of wave packets.},
\be\label{MfFromM}
{\bf M}(f;b')=\int_0^\infty\ud b_0\rho(b_0){\bf M}(b_0,b') \;,
\ee
where ${\bf M}(b_0,b')$ is determined by~\eqref{Mintegrobprime}. Note that even if we were only interested in a sharply peaked wave packet for only one value of $b_0=B$, to solve~\eqref{Mintegrobprime} we anyway need to consider ${\bf M}(b_0,b')$ for $0<b_0<B$. Thus, once we have solved~\eqref{Mintegrobprime} for a sharply peaked wave packet, there is very little extra work to perform the integral in~\eqref{MfFromM} for various types of wave packets. In other words, all the really nontrivial stuff is included in ${\bf M}(b_0,b')$, which is obtained without having to choose spin states or wave packets. We can therefore think of ${\bf M}(b_0,b')$ as a Green's function.   

Partly in order to compare with kinetic equations, we will rewrite this in terms of the spectrum (this differ by a factor of $b_0$ compared to the previous definition~\eqref{Sdefx})
\be
{\bf S}(b')=-\frac{\partial{\bf M}(f;b')}{\partial b'} \;.
\ee 
We begin by re-expanding as ${\bf S}=\sum_{n=0}^\infty{\bf S}^{(n)}(+\infty,b')$ with ${\bf S}^{(n)}=\mathcal{O}(\alpha^n)$. The zeroth order is
${\bf S}^{(0)}=\rho(b'){\bm1}$. We obtain higher orders from~\eqref{M1fromMLMC} and~\eqref{firstM2eq} by replacing 
\be
b_0\to kp_{\rm in}=b'+\sum_j kl_j
\ee
and writing the $q_j$ integrals in terms of $kl_j$. For example, in the last term in~\eqref{firstM2eq} we have
\be\label{preToApp}
\begin{split}
&\ud q_1\ud q_2{\bf M}_C(b_0,q_1,\sigma_1)\cdot{\bf M}_C([1-q_1]b_0,q_2,\sigma_2)\\
&\to\ud kl_1\ud kl_2\tilde{\bf M}_C(b'+kl_1+kl_2,kl_1,\sigma_1)\\
&\hspace{2cm}\cdot\tilde{\bf M}_C(b'+kl_2,kl_2,\sigma_2) \;,
\end{split}
\ee
where
\be
\tilde{\bf M}_C(kp,kl)=\frac{1}{kp}{\bf M}_C(kp,kl) \;,
\ee
where the first argument is for the momentum of the electron before emitting the photon.
In~\eqref{firstM2eq} and~\eqref{secondM2eq} we noted that ${\bf M}^{(2)}$ can be obtained by prepending ${\bf M}_L$ and ${\bf M}_C$ to ${\bf M}^{(1)}$. From~\eqref{preToApp} we see that ${\bf S}^{(2)}$ can instead be obtained by appending ${\bf M}_L$ and ${\bf M}_C$ to ${\bf S}^{(1)}$. We can do this if we also replace
\be
\int\ud\sigma_1\int_{\sigma_1}^\infty\ud\sigma_2\to\int\ud\sigma_2\int_{-\infty}^{\sigma_2}\ud\sigma_1 \;.
\ee
Higher orders can be obtained in the same way. Thus, we find
\be
\begin{split}
{\bf S}^{(n)}(\sigma,b')=&\int_{-\infty}^\sigma\ud\sigma'\int_0^\infty\ud(kl)\\
\times&\bigg\{{\bf S}^{(n-1)}(\sigma',b')\!\cdot\!\tilde{\bf M}_L(\sigma',b',kl)\\
&+{\bf S}^{(n-1)}(\sigma',b'+kl)\!\cdot\!\tilde{\bf M}_C(\sigma',b'+kl,kl)\bigg\} \;,
\end{split}
\ee
where
\be
\tilde{\bf M}_L(kp,kl)=\frac{1}{kp}\theta(kp-kl){\bf M}_L \;.
\ee
Summing over $n$ and differentiating with respect to $\sigma$ gives
\be\label{integroSf}
\begin{split}
\frac{\partial{\bf S}}{\partial\sigma}(b')=\int_0^\infty\ud kl&\bigg\{{\bf S}(b')\!\cdot\!\tilde{\bf M}_L(b',kl)\\
&+{\bf S}(b'+kl)\!\cdot\!\tilde{\bf M}_C(b'+kl,kl)\bigg\} \;,
\end{split}
\ee
with initial condition
\be\label{initialSf}
{\bf S}(\sigma=-\infty,b')={\bf S}^{(0)}=\rho(b'){\bf 1} \;.
\ee
If one is only interested in a definite initial Stokes vector, then one can project ${\bf N}_0\cdot\eqref{integroSf}$ before integrating and solve for ${\bf N}_0\cdot{\bf S}$ with initial condition ${\bf N}_0\cdot{\bf S}(-\infty)=\rho{\bf N}_0$. On the other hand, $\eqref{integroSf}\cdot{\bf N}_f$ does not give an equation for ${\bf S}\cdot{\bf N}_f$. For~\eqref{Mintegrobprime} we can instead project, before integration, with ${\bf N}_f$ but not with ${\bf N}_0$.
  
In contrast to the equations for cumulative function ${\bf M}$ and the moments, see e.g.~\eqref{Mintegrobprime}, which we integrate backwards in lightfront time from $\sigma=+\infty$ to $-\infty$, \eqref{integroSf} is integrated forwards from $\sigma=-\infty$ to $+\infty$. 
We can understand this difference as follows. For ${\bf S}(f;b')$ the earliest step is special due to the appearance of the initial wave packet, while the subsequent steps are obtained recursively by appending ${\bf M}_L$ and ${\bf M}_C$, which gives an equation that should be integrated forwards in $\sigma$. For the moments $\tilde{\bf M}(n,b_0)$ it is instead the last step that is special, because there we have an additional factor of $kP^n$ in the integral for the final electron, while the preceding steps are obtained recursively by prepending ${\bf M}_L$ and ${\bf M}_C$ and there is no nontrivial distribution in the first step, which gives equations that should be integrated backwards in $\sigma$. 

\begin{figure}
\includegraphics[width=\linewidth]{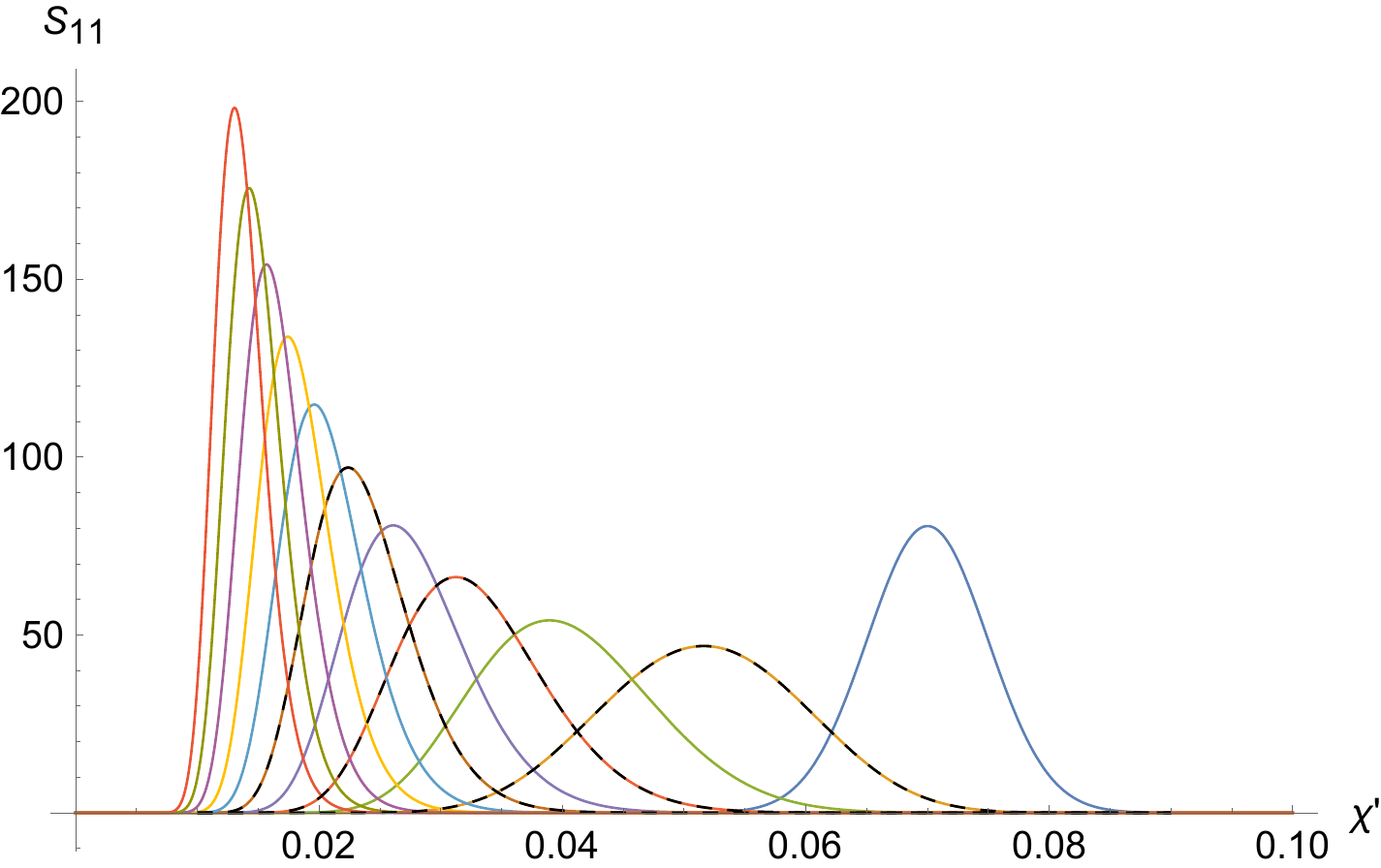}
\caption{Spectrum for a Gaussian wave packet~\eqref{gaussPacket} and a constant field, with ${\bf N}_0={\bf N}_f=\{1,0\}$. Solid lines are obtained by solving~\eqref{integroSf} with initial condition in~\eqref{initialSf}. The rightmost curve is the initial wave packet, and the other curves correspond to $T=10,20,\dots100$. The black dashed lines for $T=10,30,50$ have been obtained using~\eqref{MfFromM}.}
\label{wavePacketFig11}
\end{figure}

\begin{figure}
\includegraphics[width=\linewidth]{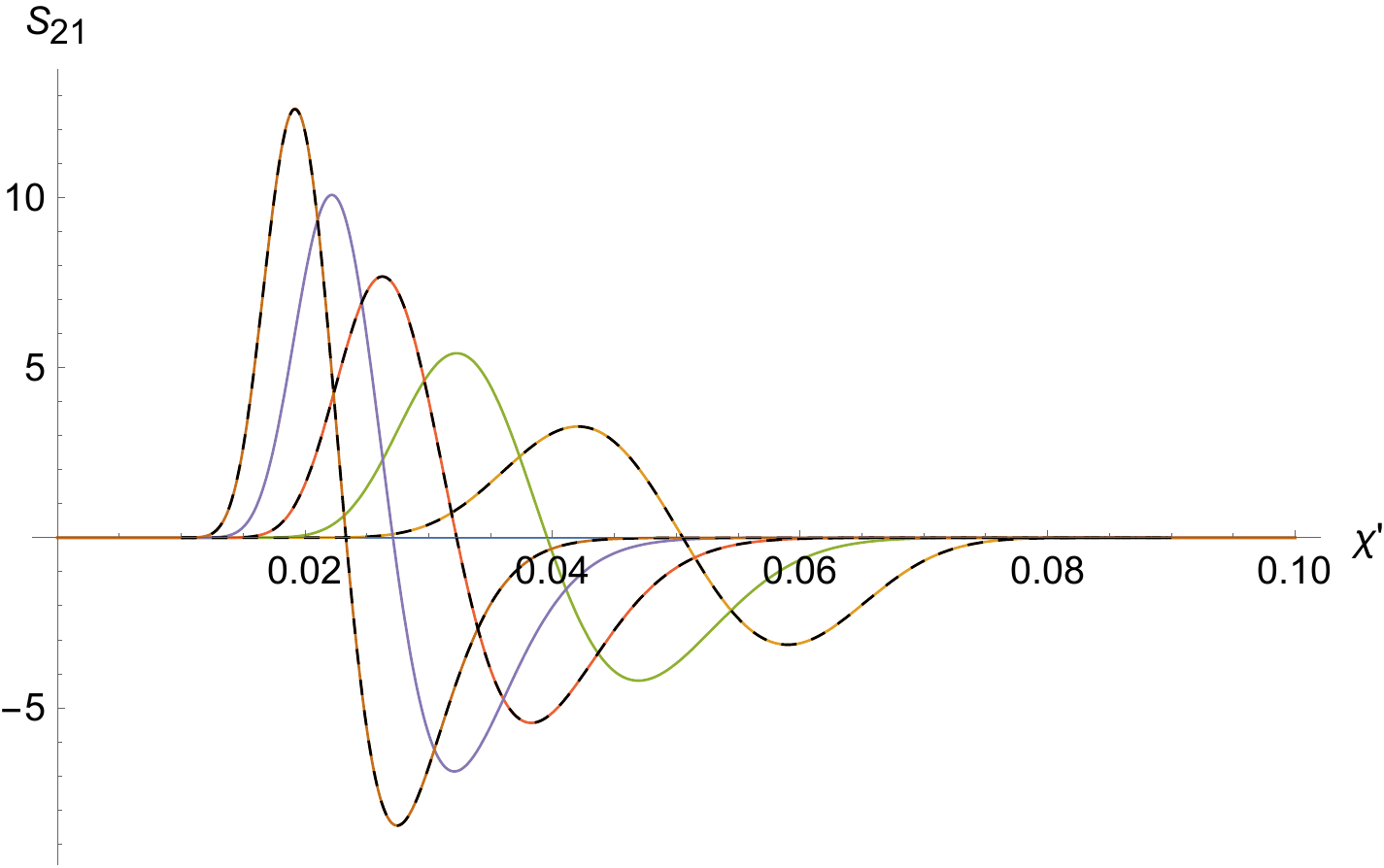}
\caption{Same as in Fig.~\ref{wavePacketFig11} but with ${\bf N}_0=\{0,1\}$ to see the dependence on the initial spin.}
\label{wavePacketFig21}
\end{figure}

\eqref{integroSf} holds as long as the field is sufficiently strong or long, i.e. not just in the LCF regime. If we restrict to the LCF regime then we can compare with kinetic equations in the literature~\cite{Shen1972,Sokolov:2010am,Neitz:2013qba,Neitz:2014hla,NielPRE2018,Elkina:2010up,Seipt:2023bcw}. If we replace
\be
{\bf S}\cdot\tilde{\bf M}_{C,L}\to
(\{1,{\bf 0}\}\!\cdot\!{\bf S}\!\cdot\!\{1,{\bf 0}\})
(\{1,{\bf 0}\}\!\cdot\!\tilde{\bf M}_{C,L}\!\cdot\!\{1,{\bf 0}\})
\ee
and identify $\{1,{\bf 0}\}\!\cdot\!{\bf S}\!\cdot\!\{1,{\bf 0}\}$ with an electron bunch (i.e. multi-particle) distribution, then we find perfect agreement with Eq.~(2) in~\cite{Neitz:2013qba}. Spin effects have recently been included in kinetic equations in~\cite{Seipt:2023bcw} using the LCF Mueller matrices from~\cite{Torgrimsson:2020gws}.

As an example we have considered a Gaussian wave packet\footnote{Strictly speaking, $\rho$ should be identically zero for $\chi<0$, but this Gaussian is anyway exponentially small there.},
\be\label{gaussPacket}
\rho(\chi)=\frac{1}{\sqrt{\pi}\sigma}\exp\left[-\frac{(\chi-\chi_0)^2}{\sigma^2}\right]
\ee
with $\chi_0=0.07$ and $\sigma=0.007$. These values have been chosen so that the wave packet is contained in $0<\chi<0.1$, so that we can compare with the results we obtained for Fig.~\ref{Schi0p1fig}. Figs.~\ref{wavePacketFig11}  and~\ref{wavePacketFig21} show perfect agreement between the results obtained from~\eqref{integroSf} and the ``Green's function approach''~\eqref{MfFromM}.

\section{Conclusions}

We have derived recursive and integrodifferential matrix equations for the momentum spectrum for electrons in a plane wave. This is formulated in terms of what can be thought of as a Green's function, ${\bf M}(b_0,b')$, which gives a complete description of RR and spin transition. After ${\bf M}(b_0,b')$ has been calculated one obtains the result for particular initial and final electron states by projecting with the initial and final Stokes vectors for the spin, ${\bf N}_0\cdot{\bf M}\cdot{\bf N}_f$, and integrating over the initial longitudinal momentum, $b_0$, weighted by the absolute square of the wave packet, as in~\eqref{MfFromM}.    

Due to the singularity at $b'=b_0$ during early ligthfront times, we have found it convenient to work with the cumulative distribution function, ${\bf M}(b_0,b')$, rather than the spectrum itself. The spectrum is then obtained at the end of the calculation by differentiation, $\partial_{b'}{\bf M}(b_0,b')$. 

By integrating the resulting spectrum we find agreement with moments calculated using a generalization of the approach in~\cite{Torgrimsson:2021wcj,Torgrimsson:2021zob}, where we considered the zeroth and first order moments. With the second moment we can calculate the standard deviation, and we find that it scales as $\sqrt{\chi}$ and is therefore relatively large even for small $\chi$.   

We have derived a low energy expansion of the spectrum for a constant field and found that it takes the form of a Gaussian multiplied by a power series. To zeroth order this tells us how the momentum is distributed around the solution to LL~\cite{exactSolLL}. We find that the expansion parameter is proportional to $\sqrt{\chi}$, so even for $\chi=0.01$ we need to go beyond the leading order and sum the first 3 or 4 orders in the $\chi\ll1$ expansion in order to obtain a precise approximation.

For the numerical results of this paper, we have chosen a constant field. We do not actually expect it to be more difficult to solve~\eqref{integroDiffx} in the LCF approximation for a nonconstant field compared to~\eqref{integroDiffConst} for a constant field. Indeed, both equations are on the form $\partial{\bf M}/\partial t={\bf F}$, where $t=\sigma$ or $t=T$, and whether ${\bf F}$ only depends on $t$ via ${\bf M}$, as in the constant field case, or also has an explicit dependence on $t$, as in the nonconstant case, we could still use the same method (e.g. the midpoint method) when integrating over $t$. In fact, we did so in~\cite{Torgrimsson:2022ndq} for the all-order probability of trident. 
The issue is instead that for the spectrum we have an additional parameter, $x$, compared to the equations in~\cite{Torgrimsson:2021wcj,Torgrimsson:2021zob,Torgrimsson:2022ndq}, so it takes much longer time to compute an interpolation function of ${\bf M}(\chi,x)$ at each point in ``time''. While this is an issue for both~\eqref{integroDiffx} and~\eqref{integroDiffConst}, the ``time'' parameter in~\eqref{integroDiffConst} is a physically relevant parameter, so when we integrate from $T=0$ to e.g. $T=100$, all the intermediate time steps give physical results. In contrast, for~\eqref{integroDiffx} we integrate from $\sigma=+\infty$ to $\sigma=-\infty$, but it is only the final result at $\sigma=-\infty$ that gives something physical.   
Moreover, in order to check the numerical results by comparing with analytical approximations, we have to generalize the low-energy approximations to nonconstant fields. We expect the approximations for a constant field to serve as a useful guide for generalizing to nonconstant fields.
 Thus, we leave a numerical study of different nonconstant fields and generalization of the approximation of the momentum spectrum for future studies.  

Moreover, our general methods are not restricted to the LCF regime, but work as long as the field is sufficiently strong or long. So, one could use for example a locally monochromatic field approximation, which we also leave for future studies.

\acknowledgements

G. T. is supported by the Swedish Research Council, Contract No. 2020-04327.

\end{document}